\documentclass[aps,12pt,a4paper,showpacs,byrevtex]{revtex4}
\usepackage{amsmath}
\usepackage{bm}
\usepackage{amssymb}
\newcommand{\ri}{\mathrm i}
\newcommand{\e}{\mathrm e}
\newcommand{\rd}{\mathrm d}

\begin{document}
\baselineskip 15pt

\title{Time evolution, cyclic solutions and geometric phases for
the generalized time-dependent harmonic oscillator}\thanks{published
in J. Phys. A {\textbf 37} (2004) 1345-1371.}



\author{Qiong-Gui Lin}
\email[]{qg_lin@163.net}


\affiliation{China Center of Advanced Science and Technology (World
Laboratory), P. O. Box 8730, Beijing 100080, People's Republic of
China}
\thanks{not for correspondence}
\affiliation{Department of Physics, Sun Yat-Sen University, Guangzhou
510275, People's  Republic of China}


\begin{abstract}\baselineskip 15pt
{The generalized time-dependent harmonic oscillator is studied.
Though several approaches to the solution of this model have been
available, yet a new approach is presented here, which is very
suitable for the study of cyclic solutions and geometric phases. In
this approach, finding the time evolution operator for the
Schr\"odinger equation is reduced to solving an ordinary differential
equation for a c-number vector which moves on a hyperboloid in a
three-dimensional space. Cyclic solutions do not exist for all time
intervals. A necessary and sufficient condition for the existence of
cyclic solutions is given. There may exist some particular time
interval in which all solutions with definite parity, or even all
solutions, are cyclic. Criterions for the appearance of such cases
are given. The known relation that the nonadiabatic geometric phase
for a cyclic solution is proportional to the classical Hannay angle
is reestablished. However, this is valid only for special cyclic
solutions. For more general ones, the nonadiabatic geometric phase
may contain an extra term. Several cases with relatively simple
Hamiltonians are solved and discussed in detail. Cyclic solutions
exist in most cases. The pattern of the motion, say, finite or
infinite, can not be simply determined by the nature of the
Hamiltonian (elliptic or hyperbolic, etc.). For a Hamiltonian with a
definite nature, the motion can changes from one pattern to another,
that is, some kind of phase transition may occur, if some parameter
in the Hamiltonian goes through some critical value.}
\end{abstract}

\maketitle



\section{Introduction}\label{s1}

The harmonic oscillator is one of the most familiar models in
physics. It is widely used in various fields, both classically and
quantum-mechanically. Several physical problems can be described by
the harmonic oscillator with time-dependent parameters
\cite{natur,paul}. More generally, such time-dependent parameters may
describe approximately the interaction of the harmonic oscillator
with some external degrees of freedom. Therefore, the time-dependent
harmonic oscillator has been the subject of many theoretical
researches over decades \cite{lewis1,lewis2,gerry,selez}. Since the
discovery of the geometric phase
\cite{berry,simon,aha,sam,wu-li,jordan,resource,li-book}, the model
has attracted more attention
\cite{chat,wang-sj,ji,lewis3,ge,liu,wang-xb,fuent}, because it is
simple and serves as a good example for the study of geometric
phases, just like the case of particles with spin and magnetic moment
moving in time-dependent magnetic fields
\cite{wang-pra,wagh1,wagh2,fer,fer-pla,layton,gao,ni,zhang,%
zhu00,pra01,jpa01,jpa02,jpa03,ni-book}. The two models are similar in
that they involve similar Lie algebras. In fact, the Hamiltonian for
spin in a magnetic field is an element of the SO(3) algebra while
that for the time-dependent harmonic oscillator is one of the SO(2,1)
algebra. However, the time-dependent harmonic oscillator is of more
interest since it has a classical counterpart, and the quantum motion
can be compared with the classical one. For example, the relation
between the geometric phase and the classical Hannay angle
\cite{hannay1,berry1,ber-han} is an interesting subject.

There exist mainly two approaches to the solution of the
time-dependent harmonic oscillator. The main point of the first
approach \cite{lewis1,lewis2} is to find an invariant operator and
its eigenstates. The second approach employs time-dependent unitary
transformations \cite{selez,wang-sj}. Here we develop another
approach to the solution of the Schr\"odinger equation. It is a
further development of the approach previously used for spin moving
in time-dependent magnetic fields \cite{jpa03}. First we find an
invariant operator, and then go further to obtain the time evolution
operator. It should be remarked that our method of finding the
invariant operator is rather different from that in the first
approach. In that approach the problem is reduced to solving a
nonlinear differential equation. In our approach, it is reduced to
solving a linear differential equation for a three-component c-number
vector, and thus is simpler. On the other hand, the time-dependent
unitary transformation approach in Ref. \cite{wang-sj} seems even
simpler than ours, but our approach can be easily generalized to
other systems where the Hamiltonian is an element of a more
complicated Lie algebra. However, the main advantage of our approach
is that it is very suitable for the study of cyclic solutions and
geometric phases.

In the literature there exists some argument that cyclic solutions
are available for any time interval, say, $[0,\tau]$ where $\tau$ is
arbitrary, because one can always choose the eigenstates of
$U(\tau)$, the time evolution operator at $\tau$, as initial
conditions at $t=0$. This is true. However, the problem is that, for
the time-dependent harmonic oscillator, $U(\tau)$ may have no
normalizable eigenstate. This is different from the case of spin,
where no problem of normalizability has to be worried. Obviously,
normalizable states are physically more interesting than
nonnormalizable ones. Moreover, if one is interested in the geometric
phases, it seems still not clear how to define it for the
nonnormalizable cyclic solutions. The difficulty lies in the
definition of the dynamical phase. If only normalizable states are to
be considered, then one should be able to tell whether there are
cyclic solutions for a given $\tau$. This is not a trivial task even
when the time evolution operator is explicitly available. In our
approach, however, the problem can be solved naturally. We will give
a necessary and sufficient condition for the existence of cyclic
solutions in an arbitrarily given time interval.

It has been shown by several authors that the nonadiabatic geometric
phase for a quantum cyclic state is equal to $-(n+1/2)$ times the
classical Hannay angle \cite{ge,liu,wang-xb}. However, it seems not
very clear under what restriction on the initial condition is this
relation valid, or whether modification to this relation is needed in
some case where cyclic solutions exist for less restricted initial
conditions. We will reestablish the above relation and show that it
is valid only for cyclic solutions with special initial conditions.
There exist several cases where more general cyclic solutions exist
in some particular time interval. Among these cases two are of
special interest. In one of the two cases all solutions are cyclic,
and in the other case all solutions with definite parity are cyclic.
We will give criteria for the appearance of such cases. In all these
cases, the nonadiabatic geometric phase contains in general an extra
term which depends on the initial condition, in addition to the one
proportional to the classical Hannay angle. Similar situations have
been found in other systems \cite{jpa02,jpa03,pla02}.

A normalizable state can be regarded as a wave packet in the
configuration space. The motion of a wave packet can be roughly
described by the change of its position and width. For the
time-dependent harmonic oscillator, we will show that if the position
of a wave packet is confined in a finite region, then its width also
remains finite, and the contrary is also true. A problem that seems
not clear concerns the relation between the pattern of the motion,
say, finite or infinite, and the nature of the Hamiltonian, that is,
elliptic, hyperbolic, or critical (see Sec. \ref{s2}). In simple
cases where the Hamiltonian is time independent (or the time
dependence lies only in an overall factor), elliptic Hamiltonian
leads to finite motion and the other ones leads to infinite motion.
However, if the Hamiltonian is time dependent, the situation is
complicated. We will see that for a Hamiltonian with a definite
nature, say, elliptic, the motion of the wave packet may exhibit
different patterns if some parameter in the Hamiltonian takes
different values. In particular, when the parameter goes through some
critical value, some kind of phase transition occurs, that is, the
motion changes from one pattern to another. At the critical value,
the motion has an independent pattern.

This paper is organized as follows. In Sec. \ref{s2} we develop some
mathematical formulas that will be used in the subsequent sections.
In Sec. \ref{s3} a new method is presented to derive the time
evolution operator for the Schr\"odinger equation. In Sec. \ref{s4} a
necessary and sufficient condition for the existence of cyclic
solutions in an arbitrarily given time interval $[0,\tau]$ is given,
and the known relation between the nonadiabatic geometric phase and
the classical Hannay angle is reestablished. In Sec. \ref{s5} we
study several cases where more cyclic solutions are available, and
give criterion for the appearance of such cases. A modification to
the above relation between the nonadiabatic geometric phase and the
classical Hannay angle are also discussed in this section. In Sec.
\ref{s6} we study several examples where explicit results are
available. The existence of cyclic solutions are discussed in detail,
and it is seen that they exist in most of the examples. The evolution
of normalizable states are also studied from the point of view of
wave packets. Various patterns of motion are revealed in these
examples, and phase transition is explicitly observed. Sec. \ref{s7}
is devoted to a brief summary. In the appendix we briefly discuss how
to extend our formalism to a more general system.

\section{The model and some mathematics}\label{s2}

The time-dependent harmonic oscillator is described by the
Schr\"odinger equation
\begin{subequations}\label{1}
\begin{equation}\label{1a}
\ri\hbar\partial_t\psi(t)=H(t)\psi(t)
\end{equation}
with the following Hamiltonian.
\begin{equation}\label{1b}
H(t)=\omega(t)\bm K\cdot\bm n^g(t),
\end{equation}
\end{subequations}
where $\omega(t)$ is a time-dependent frequency parameter, $\bm
n(t)=\bm(n_1(t),n_2(t),n_3(t)\bm)$ is a time-dependent c-number
vector and $\bm K=(K_1,K_2,K_3)$ is an operator vector defined below.
In this paper, any vector $\bm a=(a_1,a_2,a_3)$ has an associated
vector $\bm a^g=(-a_1,-a_2,a_3)$, and scalar product between two
vectors $\bm a$ and $\bm b$ always appears as $(\bm a, \bm b)=\bm a
\cdot\bm b^g=\bm a^g\cdot\bm b=a_3 b_3-a_1 b_1-a_2 b_2$. In matrix
form these are $a^g=ga$ and $(\bm a, \bm b)=a^{\mathrm t} g
b=b^{\mathrm t} g a$ where $a$ and $b$ are now column vectors and the
superscript t denote transposition, and $g=\mathrm{diag}(-1,-1,1)$.
In the following the square of a vector $\bm a^2$ always stands for
$\bm a\cdot\bm a^g$ rather than $\bm a\cdot\bm a$. In the above
Hamiltonian we always choose $\omega$ such that $\bm n^2$ equals 1,
$-1$ or 0. Thus among the components of $\bm n$ only two are
independent, and the Hamiltonian involves three independent
parameters. The vector $\bm K$ is defined as
\begin{subequations}\label{2}
\begin{equation}\label{2a}
K_1=\frac12\left(\mu_0\omega_0 X^2-{P^2\over \mu_0\omega_0}\right),
\quad K_2=-\frac12(XP+PX),\quad K_3=\frac12 \left(\mu_0\omega_0
X^2+{P^2\over \mu_0\omega_0}\right),
\end{equation}
where $X$ is the coordinate and $P$ the momentum, satisfying
$[X,P]=\ri\hbar$, $\mu_0$ is the mass of the particle, and $\omega_0$
is some constant frequency parameter. For simplicity in notations, we
define $x=\sqrt{\mu_0\omega_0}X$ and $p=P/\sqrt{\mu_0\omega_0}$,
which still satisfy the commutation relation $[x, p]=\ri\hbar$, then
the above expressions become
\begin{equation}\label{2b}
K_1=\textstyle\frac12(x^2-p^2), \quad
K_2=-\textstyle\frac12(xp+px),\quad K_3=\textstyle \frac12 (x^2+p^2).
\end{equation}
\end{subequations}
Note that both $x$ and $p$ have the same dimensionality as
$\sqrt\hbar$, and $\bm K$ has the same dimensionality as $\hbar$. The
components of $\bm K$ satisfies the commutation relation
\begin{equation}\label{3}
[K_i,K_j]=-2\ri\hbar\epsilon_{ijk}K_k^g.
\end{equation}
An equivalent form is $[K_i^g,K_j^g]=-2\ri\hbar\epsilon_{ijk}K_k$.
This is an SO(2,1) algebra. The Hamiltonian is said to be elliptic,
hyperbolic or critical if $\bm n^2$ equals 1, $-1$ or 0,
respectively.

To solve the Schr\"odinger equation we need some operator formulas.
We will briefly derive them here. Let
\begin{equation}\label{4}
\bm F(\xi)=Q(\xi, \bm b)\bm K Q^\dag(\xi,\bm b), \quad Q(\xi, \bm b)=
\exp(-\textstyle\frac12\ri\hbar^{-1}\xi \bm K\cdot\bm b^g),
\end{equation}
where $\xi$ is a real parameter and $\bm b$ a real vector independent
of $\xi$. This is a unitary transformation of $\bm K$. Using Eq.
(\ref{3}) it can be shown that
\begin{equation}\label{5}
\bm F'(\xi)=\bm b^g\times\bm F^g(\xi),
\end{equation}
where the prime indicates derivative with respect to $\xi$. An
equivalent equation is $\bm F^{g\prime}(\xi)=\bm b \times \bm
F(\xi)$, which is convenient in obtaining
\begin{equation}\label{6}
\bm F''(\xi)+\bm b^2 \bm F(\xi)=[\bm F(\xi)\cdot \bm b^g]\bm b.
\end{equation}
However, Eq. (\ref{5}) leads to $\bm F'(\xi)\cdot \bm b^g=0$, and
thus $\bm F(\xi)\cdot \bm b^g=\bm F(0)\cdot \bm b^g=\bm K\cdot \bm
b^g$. Therefore Eq. (\ref{6}) is simplified as
\begin{subequations}\label{7}
\begin{equation}\label{7a}
\bm F''(\xi)+\bm b^2 \bm F(\xi)=(\bm K\cdot \bm b^g)\bm b.
\end{equation}
This is a simple equation. With the initial condition
\begin{equation}\label{7b}
\bm F(0)=\bm K,\quad \bm F'(0)=\bm b^g\times\bm K^g,
\end{equation}
\end{subequations}
the solution is easily fixed. The results are listed below.

If $\bm b^2=1$, we denote $\bm b$ by $\bm b_+$, and have
\begin{equation}\label{8}
Q(\xi, \bm b_+)\bm K Q^\dag(\xi,\bm b_+)=[\bm K - (\bm K\cdot \bm
b_+^g)\bm b_+]\cos\xi+ \bm b_+^g \times\bm K^g \sin\xi+(\bm K\cdot
\bm b_+^g)\bm b_+.
\end{equation}
When $\xi=2N \pi$ where $N$ is an integer, we have $Q(2N \pi, \bm
b_+)\bm K Q^\dag(2N \pi,\bm b_+)=\bm K$, or $Q(2N \pi, \bm b_+)\bm
K=\bm K Q(2N \pi,\bm b_+)$. That is, $Q(2N \pi, \bm b_+)$, and in
particular, $\exp(\ri\hbar^{-1} N\pi K_3)$, commutes with $\bm K$.
However, this does not mean that $Q(2N \pi, \bm b_+)$ is a c-number,
since a c-number must commute with $x$ and $p$. But we will see below
that $Q(4N \pi, \bm b_+)$ is indeed a c-number.

If $\bm b^2=-1$, we denote $\bm b$ by $\bm b_-$, and have
\begin{equation}\label{9}
Q(\xi, \bm b_-)\bm K Q^\dag(\xi,\bm b_-)=[\bm K + (\bm K\cdot \bm
b_-^g)\bm b_-]\cosh\xi+ \bm b_-^g \times\bm K^g \sinh\xi-(\bm K\cdot
\bm b_-^g)\bm b_-.
\end{equation}
When $\bm b_-=(-\sin\phi,\cos\phi,0)$, the result is useful in the
following sections. In this case we denote $\bm b_-$ by $\bm b_\phi$
and $Q(\xi, \bm b_-)$ by $Q(\xi, \phi)$:
\begin{equation}\label{10}
Q(\xi, \phi)=\exp(-\textstyle\frac12\ri\hbar^{-1}\xi \bm K\cdot\bm
b^g_\phi) =\exp[-\textstyle\frac12\ri\hbar^{-1} \xi(K_1\sin\phi
-K_2\cos\phi)].
\end{equation}
In particular we write down
\begin{equation}\label{11}
Q(\xi, \phi) K_3 Q^\dag(\xi,\phi)=K_3\cosh\xi
-K_1\sinh\xi\cos\phi-K_2\sinh\xi\sin\phi.
\end{equation}

If $\bm b^2=0$, we denote $\bm b$ by $\bm b_0$, and have
\begin{equation}\label{12}
Q(\xi, \bm b_0)\bm K Q^\dag(\xi,\bm b_0)=\bm K + \xi \bm b_0^g
\times\bm K^g +\textstyle\frac12\xi^2(\bm K\cdot \bm b_0^g)\bm b_0.
\end{equation}

For a c-number vector $\bm a$, one can define a transformed vector
$\bm A$ by
\begin{equation}\label{13}
\bm K\cdot\bm A=Q(\xi, \bm b)\bm K\cdot\bm a Q^\dag(\xi,\bm b).
\end{equation}
It can be shown by straightforward calculations that $\bm A^2 =\bm
a^2$, regardless of the values of $\xi$ and $\bm b$. Thus the
elements of the SO(2,1) algebra, $\bm K\cdot\bm a$, are distinguished
into three classes, characterized by whether $\bm a^2$ is positive,
negative or zero. They cannot be connected by any one of the above
unitary transformations. For example, $K_3$ cannot be transformed to
$K_1$ and vice versa. Geometrically, the surface defined by $\bm
a^2=0$ is a cone in the $\bm a$ space. Therefore, the three classes
of elements are characterized by whether $\bm a$ is inside, outside
or on the cone.

Similarly, one can calculate the unitary transformation of $x$ and
$p$. Here we only write down the result for the case with $\bm b^2=1$
(the subscript $+$ in the components of $\bm b_+$ is omitted).
\begin{subequations}\label{14}
\begin{equation}\label{14a}
Q(\xi, \bm b_+) x Q^\dag(\xi,\bm b_+)= x\left(\cos{\xi\over2}
-b_{2}\sin{\xi\over2}\right) -p(b_{3}+b_{1})\sin{\xi\over2},
\end{equation}
\begin{equation}\label{14b}
Q(\xi, \bm b_+) p Q^\dag(\xi,\bm b_+)= x(b_{3}-b_{1}) \sin{\xi\over2}
+p\left(\cos{\xi\over2} +b_{2} \sin{\xi\over2}\right).
\end{equation}
\end{subequations}
Here we see that $Q(4N \pi, \bm b_+)$ commutes with $x$ and $p$. Thus
it must be a c-number, as mentioned above.

The final point of this section concerns the eigenvalues and
eigenstates of the operator $\bm K$. We are only interested in
normalizable, or bound states. $K_1$ and $K_2$ do not have
normalizable eigenstates. The eigenvalues of $K_3$ are $k_n\hbar$
where $k_n=n+1/2$ and $n=0,1,2,\ldots$. The corresponding eigenstates
will be denoted by $\psi_n$. In the coordinate representation,
$\psi_n(x)=N_n \exp(-x^2/2\hbar)H_n(x/\sqrt\hbar)$, where $H_n$ are
Hermite polynomials and $N_n=(1/2^n n!\sqrt{\pi\hbar})^{1/2}$. If a
vector $\bm e$ satisfies $\bm e^2>0$, then $\bm K\cdot \bm e^g$ has
normalizable eigenstates. Without loss of generality, we take $\bm
e^2=1$ and $e_{3}>0$, then $\bm e$ can be written as
\begin{equation}\label{15}
\bm e=(\sinh\xi\cos\phi,\sinh\xi\sin\phi,\cosh\xi).
\end{equation}
According to Eq. (\ref{11}), $\bm K\cdot \bm e^g=Q(\xi, \phi) K_3
Q^\dag(\xi,\phi)$. Therefore the eigenvalues of $\bm K\cdot \bm e^g$
are still $k_n\hbar$, and the corresponding eigenstates are
\begin{equation}\label{16}
\psi_n^{\bm e}=Q(\xi, \phi)\psi_n
=\exp(-\textstyle\frac12\ri\hbar^{-1}\xi \bm K\cdot\bm
b^g_\phi)\psi_n.
\end{equation}
This result will be employed below.

\section{Time evolution operator}\label{s3}

In this section we deal with the time evolution operator for the
Schr\"odinger equation (\ref{1}). We define a time-dependent c-number
vector $\bm e(t)$ by the differential equation
\begin{subequations}\label{17}
\begin{equation}\label{17a}
\dot {\bm e}(t)=-2\omega(t){\bm n^g}(t)\times {\bm e^g}(t)
\end{equation}
where the overdot denotes differentiation with respect to $t$, and
the initial condition
\begin{equation}\label{17b}
{\bm e}(0)={\bm e}_0,
\end{equation}
where ${\bm e}_0$ satisfies
\begin{equation}\label{17c}
\bm e_0^2=1,\quad e_{03}>0,
\end{equation}
and is otherwise arbitrary.
\end{subequations}
We would assume that $\bm n(t)$ and $\omega(t)$ varies continuously,
so that any solution ${\bm e}(t)$ is well behaved. If one solution to
this equation can be found, then the time evolution operator for the
Schr\"odinger equation can be worked out. It should be remarked that
the above equation for $\bm e(t)$ is the one satisfied by $\langle
\bm K\rangle$, the mean value of the operator vector $\bm K$ in an
arbitrary state (cf. Eq. (\ref{27}) below). More discussions can be
found in the appendix.

We take the initial state of the system to be $\psi(0)= \psi_n^{\bm
e_0}$, that is
\begin{equation}\label{18}
{\bm K}\cdot{\bm e}_0^g\psi(0)=k_n\hbar\psi(0),\quad n=0,1,2,\ldots.
\end{equation}
If $\psi$ evolves according to Eq. (\ref{1}) and $\bm e$ evolves
according to Eq. (\ref{17}), then
\begin{equation}\label{19}
{\bm K}\cdot{\bm e}^g(t)\psi(t)=k_n\hbar\psi(t)
\end{equation}
would hold at all later times. This can be easily proven by
induction.

By definition, Eq. (\ref{19}) is valid at $t=0$. We assume that it is
valid at time $t$, what we need to do is to show that it is also true
at time $t+\Delta t$ where $\Delta t$ is an infinitesimal increment
of time. In fact, using Eqs. (\ref{1}) and (\ref{17}) we have
\begin{subequations}\label{20}
\begin{equation}\label{20a}
\psi(t+\Delta t)=\psi(t)-\ri\hbar^{-1}\omega(t) {\bm K} \cdot {\bm
n^g}(t) \psi(t)\Delta t,
\end{equation}
\begin{equation}\label{20b}
{\bm e}(t+\Delta t)={\bm e}(t)-2\omega(t){\bm n^g}(t)\times{\bm
e^g}(t)\Delta t.
\end{equation}
\end{subequations}
After some simple algebra, the conclusion is achieved.

It should be remarked here that ${\bm K}\cdot{\bm e}^g(t)$ is an
invariant operator, so that it has time-independent eigenvalues.
Indeed, it is easy to show that
\begin{equation}\label{21}
\ri\hbar{\bm K}\cdot{\dot{\bm e}}^g(t)+[{\bm K}\cdot{\bm e}^g(t),
H]=0.
\end{equation}
We see that once a solution $\bm e(t)$ is found, an invariant
operator is obtained. Since $\bm e(t)$ satisfies a linear
differential equation, this method is convenient. Moreover, it can be
easily generalized to systems where the Hamiltonian is an element of
a more complicated Lie algebra. A brief discussion on this point is
given in the appendix.

From Eq. (\ref{17}) it is easy to show that $\bm e^2(t)=\bm e_0^2=1$.
This yields $|e_3(t)|\ge 1$. As is assumed, $\bm e(t)$ varies
continuously, thus $e_3(t)$ keeps its original sign at all later
times. Therefore, $\bm e(t)$ can be written in the form of Eq.
(\ref{15}), where $\xi=\xi(t)$ and $\phi=\phi(t)$, and
\begin{equation}\label{22}
\psi(t)=\exp[\ri\alpha_n(t)]Q\bm(\xi(t),\phi(t)\bm)\psi_n,
\end{equation}
where $\alpha_n(t)$ is a phase that cannot be determined by the
eigenvalue equation. However, $\alpha_n(t)$ is not arbitrary. To
satisfy the Schr\"odinger equation, it should be determined by the
other variables $\xi(t)$ and $\phi(t)$. In fact, the above equation
yields
\begin{equation}\label{23}
{\bm(}\psi(t),\psi(t+\Delta t){\bm)}=1+\ri\dot\alpha_n(t)\Delta t
+\bm (\psi_{n}, Q^\dag(\xi,\phi) \partial_t Q(\xi,\phi)\psi_n {\bm)}
\Delta t.
\end{equation}
Using the formula \cite{wilcox}
\begin{equation}\label{24}
\mathrm e^{F(t)}\partial_t \mathrm e^{-F(t)}=-\int_0^1 \mathrm
e^{\lambda F(t)}\dot F(t) e^{-\lambda F(t)}\;\rd\lambda,
\end{equation}
where $F(t)$ is any operator depending on $t$, then using Eq.
(\ref{9}), and notice that $(\psi_n,K_1\psi_n)=(\psi_n,K_2\psi_n)=0$,
we obtain
\begin{equation}\label{25}
{\bm(}\psi(t),\psi(t+\Delta t){\bm)}=1+\ri\dot\alpha_n(t)\Delta t
-\textstyle\frac12\ri k_n \dot\phi(t)[\cosh\xi(t)-1]\Delta t.
\end{equation}
On the other hand, from Eq. (\ref{20}) we have
\begin{equation}\label{26}
{\bm(}\psi(t),\psi(t+\Delta t){\bm)}=1-\ri\hbar^{-1}\omega(t){\bm
u}(t) \cdot{\bm n^g}(t)\Delta t,
\end{equation}
where
\begin{equation}\label{27}
\bm u(t)={\bm(}\psi(t),\bm K\psi(t){\bm)}.
\end{equation}
This definition will be repeatedly used below. It is easy to show
that $\bm u(t)$ satisfies the same equation as $\bm e(t)$, and for
the above initial state $\bm u(0)=k_n\hbar\bm e_0$, so we have $\bm
u(t)=k_n\hbar\bm e(t)$. Comparing the two results above and taking
this relation into account, we obtain
\begin{equation}\label{28}
\dot\alpha_n(t)=\textstyle\frac12 k_n
\dot\phi(t)[\cosh\xi(t)-1]-k_n\omega(t){\bm e}(t) \cdot{\bm n^g}(t).
\end{equation}
Therefore
\begin{equation}\label{29}
\alpha_n(t)-\alpha_n(0)=k_n\alpha(t),
\end{equation}
where
\begin{equation}\label{30}
\alpha(t)={\frac12} \int_0^t \dot\phi(t') [\cosh\xi(t')-1]\;\rd t' -
\int_0^t \omega(t'){\bm e}(t') \cdot{\bm n^g}(t') \;\rd t'.
\end{equation}
Substituting into Eq. (\ref{22}) we obtain
\begin{equation}\label{31}
\psi(t)=Q\bm(\xi(t),\phi(t)\bm) \exp[\ri\hbar^{-1}\alpha(t) K_3]
Q^\dag\bm(\xi(0),\phi(0)\bm)\psi(0).
\end{equation}
We denote the time evolution operator as $U(t)$, defined by the
equation $\psi(t)=U(t)\psi(0)$ with an arbitrary $\psi(0)$, then the
above equation is equivalent to
\begin{equation}\label{32}
U(t)\psi_n^{\bm e_0}=Q\bm(\xi(t),\phi(t)\bm)
\exp[\ri\hbar^{-1}\alpha(t) K_3]
Q^\dag\bm(\xi(0),\phi(0)\bm)\psi_n^{\bm e_0}.
\end{equation}
Now an arbitrary normalizable initial state $\psi(0)$ can be expanded
as
\begin{equation}\label{33}
\psi(0)=\sum_n c_n \psi_n^{\bm e_0}.
\end{equation}
Applying $U(t)$ to both sides of this equation, using Eq. (\ref{32}),
and noting that the operators on the right-hand side of that equation
is independent of $n$, we immediately realize that Eq. (\ref{31}) is
in fact valid for an arbitrary initial state. Thus we arrive at the
result
\begin{subequations}\label{34}
\begin{equation}\label{34a}
U(t)=\exp[-\textstyle\frac12\ri\hbar^{-1}\xi(t)\bm K\cdot\bm
b_\phi^g(t)] \exp[\ri\hbar^{-1}\alpha(t) K_3]
\exp[\textstyle\frac12\ri\hbar^{-1}\xi(0)\bm K\cdot\bm b_\phi^g(0)].
\end{equation}
Using Eq. (\ref{9}), it can be recast in the form
\begin{equation}\label{34b}
U(t)=\exp[-\textstyle\frac12\ri\hbar^{-1}\xi(t)\bm K\cdot\bm
b_\phi^g(t)] \exp[\textstyle\frac12\ri\hbar^{-1}\xi(0)\bm K\cdot\bm
b_\phi^g(0)] \exp[\ri\hbar^{-1}\alpha(t) \bm K\cdot\bm e_0^g].
\end{equation}
\end{subequations}
Eq. (\ref{34b}) is suitable for the general discussions below while
Eq. (\ref{34a}) may be more convenient for practical calculations.

Let us make some remarks on the result. First, we see that once a
solution of Eq. (\ref{17}) is found, the time evolution operator for
Eq. (\ref{1}) is available. The result depends formally on $\bm e_0$,
but $\bm e_0$ is merely an auxiliary object, hence the result must be
essentially independent of it, though it might be difficult to prove
this explicitly. On the other hand, it is the flexibility in the
choice of $\bm e_0$ that makes it convenient for the general
discussions of cyclic solutions. In practical calculations, one
should choose a solution $\bm e(t)$ that is as simple as possible
such that $U(t)$ can be easily reduced to the simplest form. Second,
the operator $U(t)$ depends not only on $\bm e(t)$, but also on the
history of it. This is obvious from Eq. (\ref{30}). Third, though
$\phi(t)$ is indefinite when $\xi(t)=0$, it is obvious that $U(t)$ is
well behaved everywhere. Fourth, by straightforward calculations it
can be shown that $\ri\hbar\partial_t U(t)=\omega(t) \bm K \cdot \bm
n^g(t) U(t)$ and $U(0)=1$, as expected. In other words, though $U(t)$
is obtained by considering the evolution of normalizable states, it
is also valid for nonnormalizable ones.

\section{Cyclic solutions and geometric phases}\label{s4}

Now we can go further to discuss cyclic solutions in any time
interval $[0,\tau]$ where $\tau$ is an arbitrarily given time. These
cyclic solutions are not necessarily cyclic in subsequent time
intervals with the same length, say, $[\tau,2\tau]$.

Since Eq. (\ref{17}) is a linear differential equation, the general
solution $\bm e(t)$ must depend on the initial vector $\bm e_0$
linearly. Thus it can be written in a matrix form
\begin{equation}\label{35}
e(t)=E(t)e_{0},
\end{equation}
where $e(t)$ and $e_0$ are column vectors, and $E(t)$ is a $3\times
3$ matrix which is obviously real. If both $\bm e^{(1)}(t)$ and $\bm
e^{(2)}(t)$ are solutions to Eq. (\ref{17}), it is easy to show that
$\bm e^{(1)}(t)\cdot\bm e^{(2)g}(t)=\bm e^{(1)}(0)\cdot\bm
e^{(2)g}(0)$. Therefore the matrix $E(t)$ satisfies
\begin{equation}\label{36}
E^{\mathrm t}(t)gE(t)=g.
\end{equation}
This yields $\det E(t)=\pm1$. As is assumed, $E(t)$ varies
continuously, and $\det E(0)=1$, so that $\det E(t)=1$. Therefore the
product of the three eigenvalues of $E(t)$ must be 1, and none can be
zero. Now if $e_\sigma$ is an eigenvector with eigenvalue $\sigma$,
that is, $E e_\sigma=\sigma e_\sigma$, it can be easily shown that
$E^{\mathrm t}(g e_\sigma)=\sigma^{-1}(g e_\sigma)$. This means that
$\sigma^{-1}$ is an eigenvalue of $E^{\mathrm t}$, and thus an
eigenvalue of $E$. Therefore the eigenvalues of $E(t)$ should be
$\{1,\sigma(t),\sigma^{-1}(t)\}$. Since $E$ is real, $\sigma^*$ is
its eigenvalue if $\sigma$ is one. Therefore, if $\sigma(t)$ is
complex, it must be unit: $|\sigma(t)|=1$.

If $\sigma(\tau)\ne 1$ at the time $\tau$, one eigenvector
$\bm\eta(\tau)$ of the matrix $E(\tau)$ with eigenvalue $1$ can be
found, which satisfies $E(\tau) \eta(\tau) =\eta(\tau)$. It can be
taken as real. $\bm\eta^2(\tau)$ may be positive, negative or zero,
depending on $E(\tau)$ and $\tau$. First we consider the case with
\begin{equation}\label{37}
\bm\eta^2(\tau)>0.
\end{equation}
Because $\bm\eta(\tau)$ is only determined up to a constant factor,
we can choose that constant such that $\bm\eta^2(\tau)=1$ and
$\eta_3>0$. Then we can take
\begin{equation}\label{38}
\bm e_0=\bm\eta(\tau)
\end{equation}
as the initial condition in Eq. (\ref{17}), and have $e(\tau)=E(\tau)
e_{0}=E(\tau) \eta(\tau)= \eta(\tau)=e_0$, that is
\begin{equation}\label{39}
\bm e(\tau)=\bm e_0.
\end{equation}
This means that $\xi(\tau)=\xi(0)$ and $\bm b_\phi(\tau)=\bm b_\phi
(0)$, and leads to
\begin{equation}\label{40}
U(\tau)=\exp[\ri\hbar^{-1}\alpha(\tau) \bm K\cdot\bm e_0^g].
\end{equation}
Now it is clear that with the initial condition $\psi(0)=\psi_n^{\bm
e_0}$ ($n=0,1,2,\ldots$), we have a cyclic solution in the time
interval $[0,\tau]$. More specifically, $\psi(\tau)=\mathrm
e^{\ri\delta_n}\psi(0)$, where the total phase change is
$\delta_n=k_n \alpha(\tau),\mathrm{mod}~2\pi$, with $\alpha(\tau)$
given by
\begin{equation}\label{41}
\alpha(\tau)=\frac12 \int_0^\tau \dot\phi(t) [\cosh\xi(t)-1]\;\rd t -
\int_0^\tau \omega(t){\bm e}(t) \cdot{\bm n^g}(t) \;\rd t.
\end{equation}
For the present state, $\bm u(t)=k_n\hbar\bm e(t)$, so the dynamic
phase $\beta_n=-\hbar^{-1}\int_0^\tau \langle H(t)\rangle\;\mathrm d
t$ turns out to be
\begin{equation}\label{42}
\beta_n=-k_n\int_0^\tau \omega(t)\bm e(t)\cdot \bm n^g(t)\;\rd t.
\end{equation}
Therefore the nonadiabatic geometric phase
$\gamma_n=\delta_n-\beta_n$ is given by
\begin{equation}\label{43}
\gamma_n=-k_n\Delta\theta_{\mathrm g},\quad \mathrm{mod}~2\pi,
\end{equation}
where
\begin{equation}\label{44}
\Delta\theta_{\mathrm g}=-\frac12 \int_0^\tau \dot\phi(t)
[\cosh\xi(t)-1]\;\rd t=-\frac12\int_0^\tau{e_1(t)\dot e_2(t)-\dot
e_1(t) e_2(t)\over e_3(t)+1}\;\rd t
\end{equation}
will be shown to be the classical Hannay angle below. Because $\bm
e^2(t)=1$ and $e_3(t)>0$, $\bm e(t)$ moves on the upper sheet of a
hyperboloid. This is a basic consequence of the fact that the
Hamiltonian is an element of the SO(2,1) algebra. The above integral
can be recast in two other forms
\begin{equation}\label{45}
\Delta\theta_{\mathrm g}=\mp\frac12 \int_S {\rd S \over
\sqrt{e_1^2+e_2^2+e_3^2}} =\mp\frac12 \int_{S_{12}}{\rd S_{12}\over
\sqrt{1+e_1^2+e_2^2}},
\end{equation}
where $S$ is the surface enclosed by the closed trace of $\bm e(t)$
on the hyperboloid, and $\rd S$ the surface element; $ S_{12}$ is the
projection of $S$ on the $e_1 e_2$ plane, and $\rd S_{12}$ the area
element; the upper (lower) sign corresponds to an anticlockwise
(clockwise) trace of $\bm e(t)$. The geometric nature of the Hannay
angle is obvious from the above expression, because it depends only
on the closed trace of $\bm e(t)$, but not on the details of the
traversing process. Because of the relation (\ref{43}), the geometric
nature of the nonadiabatic geometric phase is also obvious.

Thus Eq. (\ref{37}) is a sufficient condition for the existence of
cyclic solutions. Under this condition there exist at least a
denumerable set of normalizable cyclic solutions in the time interval
$[0,\tau]$. Of course they may be trivial ones in some cases. All
phases can be expressed in terms of the vector $\bm e(t)$. The
relation between the nonadiabatic geometric phase and the Hannay
angle is reestablished.

States with initial condition other than the above ones are in
general not cyclic ones, even though in these initial states $\bm
u(0)$ points in the direction of $\bm e_0$ such that $\bm u(\tau)=\bm
u(0)$. However, if $\alpha(\tau)/\pi$ happens to be a rational
number, more cyclic solutions are available, and the above relation
between the nonadiabatic geometric phase and the Hannay angle would
need modification for these cyclic solutions. This will be discussed
in the next section.

If $\bm\eta^2(\tau)\le0$, one can still take $\bm e(0)=\bm\eta(\tau)$
as the initial condition for Eq. (\ref{17a}), and have $\bm
e(\tau)=\bm e(0)$. However, this solution cannot be used in the time
evolution operator (\ref{34}), and no similar discussions to the
above are available. In fact, there is no normalizable cyclic
solution in this case, since the condition (\ref{37}) is also a
necessary one. This is proven below.

If there exist one normalizable cyclic solution in the time interval
$[0,\tau]$, that is, $\psi(\tau)=\e^{\ri\delta}\psi(0)$, then in this
state $\bm u(\tau)=\bm u(0)$, or $u(\tau)=u(0)$ in the form of column
vectors. As pointed out before, $\bm u(t)$ satisfies the same
equation as $\bm e(t)$, thus $u(\tau)=E(\tau)u(0)$. Comparing the two
relations we obtain $E(\tau)u(0)=u(0)$. In other words, $u(0)$ is an
eigenvector of $E(\tau)$ with eigenvalue 1. The remaining point is to
show that $\bm u^2(0)>0$. Indeed, for any normalizable state
$\psi(t)$, it is not difficult to show that $\bm
u^2(t)\ge\hbar^2/4>0$, by using the Schwarz inequality.

To conclude this section let us work out the Hannay angle in terms of
$\bm e(t)$. We denote the classical coordinate by $q_{\mathrm c}$ and
momentum by $p_{\mathrm c}$, and define a vector $\bm I$ as
\begin{equation}\label{46}
I_1=\textstyle\frac12(q_{\mathrm c}^2-p_{\mathrm c}^2), \quad
I_2=-q_{\mathrm c} p_{\mathrm c}, \quad I_3=\frac12(q_{\mathrm
c}^2+p_{\mathrm c}^2).
\end{equation}
The classical Hamiltonian is now $H_{\mathrm c}=\omega(t)\bm I\cdot
\bm n^g(t)$, and the evolution of $q_{\mathrm c}$ and $p_{\mathrm c}$
are governed by the canonical equations of motion. It is easy to show
that $I=\bm I\cdot\bm e^g(t)$ is an invariant. This leads to a
quadratic equation in $q_{\mathrm c}$ and $p_{\mathrm c}$:
\begin{equation}\label{47}
(e_3+e_1)p_{\mathrm c}^2+2e_2 q_{\mathrm c} p_{\mathrm
c}+(e_3-e_1)q_{\mathrm c}^2=2I.
\end{equation}
If $\bm e^2=1$, this describes an ellipse on the $q_{\mathrm c}
p_{\mathrm c}$ plane, whose area is $2\pi I$. The ellipse changes its
shape when $\bm e(t)$ varies with time. If $\bm e(\tau)=\bm e(0)$,
the ellipse at the time $\tau$ coincide with that at the initial
time. This is a classical nonadiabatic cyclic evolution. The
$q_{\mathrm c}$ and $p_{\mathrm c}$ can be expressed in terms of $I$
and its canonical variables $\theta$ as
\begin{equation}\label{48}
q_{\mathrm c}(\theta, I, \bm e)=\sqrt{2I(e_3+e_1)}\cos\theta, \quad
p_{\mathrm c}(\theta, I, \bm e)=-\sqrt{2I\over e_3+e_1}
(e_2\cos\theta +\sin\theta).
\end{equation}
Using $\bm e^2=1$, we have $\rd p_{\mathrm c}\wedge \rd q_{\mathrm c}
=I\cos^2\theta \rd e_1\wedge\rd e_2/e_3$, and the contour average is
$\langle \rd p_{\mathrm c}\wedge \rd q_{\mathrm c} \rangle =I \rd
e_1\wedge\rd e_2/2e_3$. According to Ref. \cite{ber-han},
$\Delta\theta_{\mathrm g}= -\partial_I \int \langle \rd p_{\mathrm
c}\wedge \rd q_{\mathrm c} \rangle$, we arrive at
\begin{equation}\label{49}
\Delta\theta_{\mathrm g}=-\frac12\int_{S_{12}}{\rd e_1\wedge\rd
e_2\over e_3}.
\end{equation}
Note that $\rd e_1\wedge\rd e_2$ corresponds to $\rd S_{12}$ ($-\rd
S_{12}$) for an anticlockwise (clockwise) trace of $\bm e(t)$, and
$e_3=\sqrt{1+e_1^2+e_2^2}$, this is the same as Eq. (\ref{45}). It is
independent of $I$.

\section{More on cyclic solutions and geometric phases}\label{s5}

In the last section we have shown that Eq. (\ref{37}) is a sufficient
and necessary condition for the existence of cyclic solutions in the
time interval $[0,\tau]$. Under this condition, there exist a
denumerable set of cyclic solutions. In this section we discuss some
special cases where more general cyclic solutions are available. We
will see that the simple relation (\ref{43}) has to be modified.

Before discussing these cases, we define
\begin{equation}\label{50}
\bar x(t)=\bm(\psi(t),x\psi(t)\bm),\quad \bar
p(t)=\bm(\psi(t),p\psi(t)\bm)
\end{equation}
for any state $\psi(t)$, and study their evolution with time.
According to the Schr\"odinger equation, they satisfy the equation of
motion:
\begin{equation}\label{51}
\dot{\bar x}=\omega[n_2\bar x +(n_1+n_3)\bar p], \quad \dot{\bar
p}=\omega[(n_1-n_3)\bar x -n_2\bar p].
\end{equation}
This is the same as that for the classical variables $q_{\mathrm c}$
and $p_{\mathrm c}$, since the Hamiltonian is quadratic in $x$ and
$p$. We denote a two-component column vector $q=(\bar x,\bar
p)^{\mathrm t}$. Because the about equation is linear, we have
\begin{equation}\label{52}
q(t)=E_q(t) q(0),
\end{equation}
where $E_q(t)$ is a $2\times 2$ evolution matrix independent of
$q(0)$. Next we define a vector $\bm v(t)$ as
\begin{equation}\label{53}
v_1=\textstyle\frac12(\bar x^2-\bar p^2),\quad v_2=-\bar x\bar p,
\quad v_3=\textstyle\frac12(\bar x^2+\bar p^2).
\end{equation}
It is straightforward to show that $\bm v(t)$ satisfies the same
equation of motion as $\bm e(t)$ or $\bm u(t)$. Therefore the
evolution matrix for $\bm v(t)$ is $E(t)$. On the other hand, the
above definition can be written as
\begin{equation}\label{54}
v_i(t)=\textstyle\frac12 q^{\mathrm t}(t) J_i q(t),
\end{equation}
where
\begin{equation}\label{55}
J_1=\sigma_z, \quad J_2=-\sigma_x, \quad J_3=1,
\end{equation}
and the $\sigma$'s are Pauli matrices. Substituting Eq. (\ref{52})
into Eq. (\ref{54}), we have
\begin{equation}\label{56}
v_i(t)=\textstyle\frac12 q^{\mathrm t}(0) [E_q^{\mathrm t}(t) J_i
E_q(t)] q(0).
\end{equation}
Now any $2\times 2$ matrix can be expanded in terms of the above
$J_i$'s and $J_0=\ri \sigma_y$. Note that $J_i$ are symmetric while
$J_0$ is antisymmetric, and the matrices $E_q^{\mathrm t}(t) J_i
E_q(t)$ are symmetric, we have
\begin{equation}\label{57}
E_q^{\mathrm t}(t) J_i E_q(t)=a_{ij}(t)J_j.
\end{equation}
It is easy to show that $\mathrm{tr}(J_i J_j)=2\delta_{ij}$, and this
yields $a_{ij}(t)=\textstyle\frac12 \mathrm{tr}[E_q^{\mathrm t}(t)
J_i E_q(t)J_j]$. Substituting into Eq. (\ref{56}), we obtain $v_i(t)
=a_{ij}(t)v_j(0)$. Therefore, $E_{ij}(t)=a_{ij}(t)$, that is
\begin{equation}\label{58}
E_{ij}(t)=\textstyle\frac12 \mathrm{tr}[E_q^{\mathrm t}(t) J_i
E_q(t)J_j].
\end{equation}
In other words, if the classical equation of motion is solved, which
gives $E_q(t)$, then $E(t)$ can be obtained by simple algebraic
calculations. This indicates a relation between our formalism and
those of some previous authors, who find the time evolution operator
of the Schr\"odinger equation by solving the classical equation of
motion \cite{wang-sj,selez}. However, our formalism, where $\bm e(t)$
plays the central role, is more convenient for the discussions of
cyclic solutions and geometric phases, for example, in obtaining the
necessary and sufficient condition (\ref{37}). In practical
calculations, it is usually more convenient to solve Eq. (\ref{17})
directly than using the above relation. However, this relation is
convenient for some general discussions. For example, when
$E_q(t)=\pm 1$, it is obvious that $E_{ij}(t)=\delta_{ij}$, or
$E(t)=1$. This will be useful below.

Now we go into the main subject of this section. On the premise of
Eq. (\ref{39}), the time evolution operator is given by Eq.
(\ref{40}), where $\alpha(\tau)$ depends on the direction of $\bm
e_0$. If it happens that $\alpha(\tau)/\pi$ is a rational number,
then more cyclic solutions are available. Of special interest are the
cases where $\alpha(\tau)=2N\pi$ and $\alpha(\tau)=(2N+1)\pi$. These
will be discussed separately.

\subsection{$\alpha(\tau)=2N \pi$}\label{s5.1}

In this case $U(\tau)$ becomes a c-number. In fact, comparing Eqs.
(\ref{40}) and (\ref{4}), we find that
$U(\tau)=Q\bm(-2\alpha(\tau),\bm e_0 \bm)$. If $\alpha(\tau)$ takes
the above value, then $U(\tau)=Q(-4N \pi,\bm e_0)$, which has been
shown to be a c-number in Sec. \ref{s2}. The value of this number can
be obtained by applying $U(\tau)$ to a specific state, say,
$\psi_0^{\bm e_0}$, the ground state of $\bm{K}\cdot \bm{e}_0^g$,
which gives the result
\begin{equation}\label{59}
U(\tau)=\e^{\ri N \pi}.
\end{equation}
Several consequences can be deduced in this case.

(1) All solutions are cyclic in the time interval $[0,\tau]$,
including nonnormalizable states, though we are only interested in
normalizable ones.

(2) Let $\tilde {\bm e}_0
=(\sinh\tilde\xi_0\cos\tilde\phi_0,\sinh\tilde\xi_0\sin\tilde\phi_0,
\cosh\tilde\xi_0)$, where $\tilde\xi_0$ and $\tilde\phi_0$ are
arbitrary. One can choose $\psi(0)=\psi_0^{{\tilde{\bm e}}_0}=
Q(\tilde\xi_0, \tilde\phi_0)\psi_0$ as an initial state such that
$\bm u(0)=\hbar\tilde {\bm e}_0/2$. In the state $\psi(t)$ with the
above initial condition $\psi(0)$, we have $\bm u(t)=\hbar\tilde {\bm
e}(t)/2$, where $\tilde {\bm e}(t)$ is the solution to Eq.
(\ref{17a}) with the initial condition $\tilde {\bm e}_0$, because
${\bm u}(t)$ and $\tilde {\bm e}(t)$ satisfy the same equation. Now
\begin{equation}\label{60}
\bm u(\tau)=\bm(\psi(\tau),\bm K\psi(\tau)\bm)
=\bm(\psi(0),U^\dag(\tau)\bm K U(\tau)\psi(0)\bm) =\bm(\psi(0),\bm K
\psi(0)\bm)= \bm u(0).
\end{equation}
Therefore
\begin{equation}\label{61}
\tilde {\bm e}(\tau)={\tilde {\bm e}}_0.
\end{equation}
Because $\tilde\xi_0$ and $\tilde\phi_0$ are arbitrary, this means
that the evolution matrix $E(t)$ is a unit matrix at the time $\tau$:
\begin{equation}\label{62}
E(\tau)=1.
\end{equation}
In this case, we have obviously $\sigma(\tau)=1$, that is, all three
eigenvalues of $E(\tau)$ are 1. The inverse is not true, however. In
some cases we have three eigenvalues all equal to 1, but $E(\tau)$
cannot be diagonalized, and is of course not a unit matrix (see Sec.
\ref{s6}).

(3) Now we take ${\tilde {\bm e}}_0$, different from $\bm e_0$, as
the initial condition for Eq. (\ref{17a}), and use ${\tilde {\bm
e}}(t)$ instead of $\bm e(t)$ in Eq. (\ref{34b}), we obtain
\begin{equation}\label{63}
U(\tau)=\exp[\ri\hbar^{-1}\tilde\alpha(\tau) \bm K\cdot \tilde{\bm
e}_0^g],
\end{equation}
where $\tilde\alpha(\tau)$ is given by Eq. (\ref{41}), with $\bm
e(t)$, $\xi(t)$ and $\phi(t)$ replaced by $\tilde{\bm e}(t)$,
$\tilde\xi(t)$ and $\tilde\phi(t)$, repectively. This must be equal
to that in Eq. (\ref{59}), however. Thus we should have
$\tilde\alpha(\tau)=2 \tilde N \pi$, and $\tilde N-N$ must be an even
integer. Actually, we will show that
$\tilde\alpha(\tau)=\alpha(\tau)$, or $\tilde N=N$.

Consider two initial unit vectors $\bm e_0$ and $\tilde{\bm e}_0$,
whose difference $\delta \bm e_0=\tilde{\bm e}_0-\bm e_0$ is
infinitesimal (then $\delta \bm e_0 \cdot \bm e_0^g=0$). The
difference in $\alpha(\tau)$, according to Eq. (\ref{41}), must be
infinitesimal because the difference in $\bm e(t)$, and thus $\xi(t)$
and $\dot\phi(t)$ are all infinitesimal. Therefore $\alpha(\tau)$ and
thus $\alpha(\tau)/\pi$ are continuous functions of $\bm e_0$. Now
that $\alpha(\tau)/\pi$ can take only on even integers, an obvious
consequence is that $\alpha(\tau)$ is a constant, independent of
$\xi_0$ and $\phi_0$.

(4) Consider a cyclic solution in $[0,\tau]$ with an arbitrary
initial condition $\psi(0)$ which is normalizable. The average value
of $\bm K$ in $\psi(0)$ is denoted by $\bm u(0)$ as before. Since
$\bm u^2(0)\ge\hbar^2/4>0$, we define $u_0=\sqrt{\bm u^2(0)}/\hbar$
which is dimensionless, and introduce
\begin{equation}\label{64}
\bm e_0=\bm u(0)/\hbar u_0.
\end{equation}
Obviously, $\bm e^2_0=1$, and $e_{03}>0$ because $u_3(0)>0$, thus
this $\bm e_0$ can be used as the initial condition in Eq.
(\ref{17}). At later times, $\bm e(t)=\bm u(t)/\hbar u_0$. The
dynamic phase is
\begin{equation}\label{65}
\beta=-\hbar^{-1}\int_0^\tau \omega(t){\bm u}(t) \cdot{\bm n^g}(t)
\;\rd t.
\end{equation}
Though $\alpha(\tau)=2N \pi$ is independent of $\bm e_0$, we must
take the one given by Eq. (\ref{64}) such that the second term in Eq.
(\ref{41}) can be related to the dynamic phase above. Then
\begin{equation}\label{66}
\beta=u_0[\alpha(\tau)+\Delta\theta_{\mathrm g}] =u_0(2N \pi
+\Delta\theta_{\mathrm g}),
\end{equation}
where $\Delta\theta_{\mathrm g}$ is calculated by substituting the
above $\bm e(t)$ into Eq. (\ref{44}). Because of Eq. (\ref{59}), the
total phase change is $\delta=N \pi$, mod $2\pi$. Therefore the
geometric phase $\gamma=\delta-\beta$ turns out to be
\begin{equation}\label{67}
\gamma=-u_0\Delta\theta_{\mathrm g}-(u_0-\textstyle\frac12)2N \pi,
\quad \mathrm{mod}~2\pi.
\end{equation}
Here the first term is the familiar one, but an extra term appears,
which depends on the initial condition. It vanishes (mod $2\pi$ of
course) when $u_0-1/2$ is an integer, especially when the initial
state is an eigenstate of $\bm K\cdot \bm e_0^g$ (it cannot be an
eigenstate of $\bm K\cdot \tilde{\bm e}_0^g$ with some other
$\tilde{\bm e}_0$ since otherwise $\bm u(0)$ would point in the
direction of $\tilde{\bm e}_0$) such that $\bm u(0)=k_n\hbar\bm e_0$
and $u_0=k_n$. In the latter case it reduces to Eq. (\ref{43}) as
expected.

(5) We have seen that Eq. (\ref{39}) plus the condition
$\alpha(\tau)=2N\pi$ leads to Eq. (\ref{59}), and as a result, all
solutions are cyclic in the time interval $[0,\tau]$. If $\bm e(t)$
is complicated, however, it is not convenient to use the above
criterion because it may be difficult to calculate $\alpha(\tau)$.
Thus some other convenient criterion is of interest. Now we give a
necessary and sufficient condition for Eq. (\ref{59}):
\begin{equation}\label{68}
E_q(\tau)=1 \quad \Longleftrightarrow \quad U(\tau)=\e^{\ri N \pi}.
\end{equation}
First, suppose that $U(\tau)=\e^{\ri N \pi}$. For arbitrarily given
values $x_0$ and $p_0$, it is easy to find an initial state $\psi(0)$
such that $\bar x(0)=x_0$ and $\bar p(0)=p_0$. It is obvious that
$\bar x(\tau)=\bar x(0)$, $\bar p(\tau)=\bar p(0)$. Since $x(0)$ and
$p(0)$ are arbitrary, we obtain $E_q(\tau)=1$. Second, suppose that
$E_q(\tau)=1$. Then we have $E(\tau)=1$ as mentioned below Eq.
(\ref{58}), and $\bm e(\tau)=\bm e_0$ for any $\bm e_0$. Now we
choose $\bm e_0=(0,0,1)$, and have from Eq. (\ref{40})
$U(\tau)=\exp[\ri\hbar^{-1}\alpha(\tau) K_3]$. On the other hand,
$E_q(\tau)=1$ leads to $\bar x(\tau)=\bar x(0)$ and $\bar
p(\tau)=\bar p(0)$ for an arbitrary $\psi(0)$, that is
\begin{equation}\label{69}
\bm(\psi(0),U^\dag(\tau)xU(\tau)\psi(0)\bm)
=\bm(\psi(0),x\psi(0)\bm), \quad
\bm(\psi(0),U^\dag(\tau)pU(\tau)\psi(0)\bm)
=\bm(\psi(0),p\psi(0)\bm).
\end{equation}
Since $\psi(0)$ is arbitrary, we should have $U^\dag(\tau)xU(\tau)=x$
and $U^\dag(\tau)pU(\tau)=p$. From Eq. (\ref{14}), this is valid only
when $\alpha(\tau)=2N\pi$, which leads to $U(\tau)=\e^{\ri N \pi}$.

\subsection{$\alpha(\tau)=(2N+1) \pi$}\label{s5.2}

In this case, $U(\tau)=\exp[\ri\hbar^{-1}\alpha(\tau) \bm K\cdot\bm
e_0^g]=Q(\xi_0,\phi_0)\exp[\ri\hbar^{-1}(2N+1)\pi K_3]
Q^\dag(\xi_0,\phi_0)$, where we have used Eq. (\ref{11}) in obtaining
the second equality. As pointed out in Sec. \ref{s2},
$\exp[\ri\hbar^{-1}(2N+1)\pi K_3]$ commutes with $\bm K$, and thus
commutes with $Q(\xi_0,\phi_0)$, we have
\begin{equation}\label{70}
U(\tau)=\exp[\ri\hbar^{-1}(2N+1)\pi \bm K\cdot\bm e_0^g]
=\exp[\ri\hbar^{-1}(2N+1)\pi K_3]=\e^{\ri N\pi}\exp(\ri\hbar^{-1} \pi
K_3).
\end{equation}
Several consequences similar to those in subsection \ref{s5.1} can be
deduced in this case.

(1) All normalizable solutions with definite parity are cyclic in the
time interval $[0,\tau]$. In fact, an initial state with definite
parity can be expanded as
\begin{equation}\label{71}
\psi^{+}(0)=\sum_{n=0}^\infty a_{2n}\psi_{2n}, \quad
\psi^{-}(0)=\sum_{n=0}^\infty a_{2n+1}\psi_{2n+1},
\end{equation}
where the superscript $+$ ($-$) indicates even (odd) parity, and time
evolution does not change the parity of an initial state since $U(t)$
only involves $\bm K$ and $\bm K$ is quadratic in $x$ and $p$. It is
easy to see that
\begin{equation}\label{72}
\psi^{\pm}(\tau)=\exp(\ri\delta_\pm)\psi^{\pm}(0),
\end{equation}
where
\begin{equation}\label{73}
\delta_\pm=\pm(N+\textstyle\frac12)\pi,\quad \mathrm{mod}~2\pi.
\end{equation}

(2) Repeat the discussions of the second point in subsection
\ref{s5.1}. Though $U(\tau)$ is not a c-number now, it commutes with
$\bm K$. Thus Eq. (\ref{60}) is still valid, and so is Eq.
(\ref{62}).

(3) As before, we take ${\tilde {\bm e}}_0$, different from $\bm
e_0$, as the initial condition for Eq. (\ref{17a}), and obtain
\begin{equation}\label{74}
U(\tau)=\exp[\ri\hbar^{-1}\tilde\alpha(\tau) \bm K\cdot \tilde{\bm
e}_0^g]=Q(\tilde\xi_0,\tilde\phi_0)\exp[\ri\hbar^{-1}\tilde\alpha(\tau)
K_3] Q^\dag(\tilde\xi_0,\tilde\phi_0),
\end{equation}
This must be equal to that in Eq. (\ref{70}), however. Because
$\exp[\ri\hbar^{-1}(2N+1)\pi K_3]$ commutes with
$Q(\tilde\xi_0,\tilde\phi_0)$, we have
$\exp[\ri\hbar^{-1}\tilde\alpha(\tau) K_3]
=\exp[\ri\hbar^{-1}(2N+1)\pi K_3]$. This means that
$\tilde\alpha(\tau)=(2\tilde N+1) \pi$, and $\tilde N-N$ must be an
even integer. By arguments similar to those in subsection \ref{s5.1},
we can conclude that $\tilde\alpha(\tau)=\alpha(\tau)$.

(4) Consider a cyclic solution in $[0,\tau]$ with a normalizable
initial state $\psi(0)$ which is of definite parity. We define $\bm
e_0$ as in Eq. (\ref{64}) and use it as the initial condition in Eq.
(\ref{17}), then $\bm e(t)=\bm u(t)/\hbar u_0$. The dynamic phase is
of the form in Eq. (\ref{65}). As before, we use the above $\bm e(t)$
to calculate $\alpha(\tau)$, then the second term in $\alpha(\tau)$
can be related to the dynamic phase, and
\begin{equation}\label{75}
\beta=u_0[\alpha(\tau)+\Delta\theta_{\mathrm g}] =u_0[(2N+1) \pi
+\Delta\theta_{\mathrm g}],
\end{equation}
where $\Delta\theta_{\mathrm g}$ is calculated by substituting the
above $\bm e(t)$ into Eq. (\ref{44}). The total phase change has been
given in Eq. (\ref{73}). Therefore the geometric phase turns out to
be
\begin{equation}\label{76}
\gamma_\pm=-u_0\Delta\theta_{\mathrm g}-(u_0\mp\textstyle\frac12)
(2N+1) \pi, \quad \mathrm{mod}~2\pi.
\end{equation}
Here an extra term appears once again, which depends on the initial
condition. It vanishes (mod $2\pi$ of course) when $u_0\mp1/2$
happens to be an even integer. For example, if the initial state is
an eigenstate of $\bm K\cdot \bm e_0^g$ (then it is of definite
parity) such that $\bm u(0)=k_n\hbar\bm e_0$ and $u_0=k_n$, then the
extra term vanishes as expected.

(5) As in subsection \ref{s5.1}, we give a necessary and sufficient
condition for Eq. (\ref{70}) which may be more convenient:
\begin{equation}\label{77}
E_q(\tau)=-1 \quad \Longleftrightarrow \quad U(\tau)
=\exp[\ri\hbar^{-1}(2N+1)\pi K_3].
\end{equation}
First, suppose that $U(\tau)=\exp[\ri\hbar^{-1}(2N+1)\pi K_3]$. Using
Eq. (\ref{14}), it is easy to show that
\begin{equation}\label{78}
U^\dag(\tau)xU(\tau)=-x, \quad U^\dag(\tau)pU(\tau)=-p.
\end{equation}
For arbitrarily given values $x_0$ and $p_0$, it is easy to find an
initial state $\psi(0)$ such that $\bar x(0)=x_0$ and $\bar
p(0)=p_0$. The above equation yields $\bar x(\tau)=-\bar x(0)$, $\bar
p(\tau)=-\bar p(0)$. Since $x(0)$ and $p(0)$ are arbitrary, we obtain
$E_q(\tau)=-1$. Second, suppose that $E_q(\tau)=-1$. Then we have
$E(\tau)=1$, and $\bm e(\tau)=\bm e_0$ for any $\bm e_0$. Now we
choose $\bm e_0=(0,0,1)$, and have from Eq. (\ref{40})
$U(\tau)=\exp[\ri\hbar^{-1}\alpha(\tau) K_3]$. On the other hand,
$E_q(\tau)=-1$ leads to $\bar x(\tau)=-\bar x(0)$ and $\bar
p(\tau)=-\bar p(0)$ for an arbitrary $\psi(0)$. Thus Eq. (\ref{78})
must be true, and from Eq. (\ref{14}), we have
$\alpha(\tau)=(2N+1)\pi$, which leads to
$U(\tau)=\exp[\ri\hbar^{-1}(2N+1)\pi K_3]$.

\subsection{Other cases}\label{s5.3}

Finally we briefly discuss the case where $\alpha(\tau)/\pi$ is a
rational number other than an integer. More specifically, let
$\alpha(\tau)=r_0\pi/s_0$ where both $r_0$ and $s_0$ are integers and
prime to each other, and $s_0\ge 2$. In addition to the denumerable
set of cyclic solutions discussed before, we have also other ones in
this case. For example, the initial conditions
\begin{equation}\label{79}
\psi(0)=\sum_{s=0}^\infty a_s \psi_{2s_0 s}^{\bm e_0}
=Q(\xi_0,\phi_0)\sum_{s=0}^\infty a_s \psi_{2s_0 s}, \quad
\sum_{s=0}^\infty |a_s|^2=1
\end{equation}
gives cyclic solutions in the time interval $[0,\tau]$. In fact, it
is easy to show that $\psi(\tau)=\e^{\ri\delta}\psi(0)$ with
$\delta=\alpha(\tau)/2=r_0\pi/2s_0$. In the above initial states, we
have $\bm u(0)=\hbar u_0\bm e_0$ where $u_0=2s_0\bar s+1/2$ and $\bar
s=\sum_{s=0}^\infty s|a_s|^2$. The dynamic phase is $\beta=u_0
[\alpha(\tau)+\Delta\theta_{\mathrm g}]$. Thus the geometric phase is
\begin{equation}\label{80}
\gamma=-u_0 \Delta\theta_{\mathrm g}-2 r_0\bar s\pi, \quad
\mathrm{mod}~2\pi.
\end{equation}
We find once again that an extra term appears in the geometric phase.
This term disappears when $\bar s$ happens to be an integer,
especially when the above initial state involves only one term.

\section{Some examples with explicit results}\label{s6}

In this section we calculate some simple examples where explicit
analytic results are available. The Hamiltonian involved in these
examples may be elliptic, hyperbolic or critical. The motion of the
particle exhibits various possible patterns. A normalizable state can
be regarded as a wave packet in the configuration space. The center
of the wave packet is characterized by $\bar x$ and its velocity
characterized by $\bar p$, and their evolution is governed by the
matrix $E_q(t)$. If $E_q(t)$ is finite at all times (for example, its
elements are trigonometric functions of $t$), then the motion is said
to be finite, because both $\bar x$ and $\bar p$ will remain finite
at all times. Otherwise it is said to be infinite. The latter case
contains still different patterns, for example, $E_q(t)$ may increase
with time or may be oscillating with increasing amplitude. Other
quantities that characterize a wave packet are mainly the variances
$\Delta x=\sqrt{\langle(x-\bar x)^2\rangle}$ and $\Delta
p=\sqrt{\langle(p-\bar p)^2\rangle}$. They can be regarded as the
width of the wave packet in the coordinate and momentum space,
respectively. From Eq. (\ref{58}) we see that if $E_q(t)$ is finite
(infinite), then $E(t)$ is essentially finite (infinite) as well. We
will see that the time evolution of $(\Delta x)^2$ and $(\Delta p)^2$
are essentially governed by $E(t)$, thus the width and the position
of a wave packet exhibit similar patterns of motion in this system.
In other words, if the wave packet moves in a confined region, then
its width also remains finite, and the contrary is also true. In
fact, we can define a new vector $\bm w(t)$ by
\begin{equation}\label{81}
\bm w(t)=\bm u(t)-\bm v(t).
\end{equation}
Since both $\bm u(t)$ and $\bm v(t)$ satisfy the same equation as
$\bm e(t)$, it is obvious that $\bm w(t)$ also satisfy the same
equation, and thus its time evolution is governed by the matrix
$E(t)$. On the other hand, it is easy to show that
\begin{subequations}\label{82}
\begin{equation}\label{82a}
w_1=\textstyle\frac12[(\Delta x)^2-(\Delta p)^2], \quad
w_3=\textstyle\frac12[(\Delta x)^2+(\Delta p)^2],
\end{equation}
\begin{equation}\label{82b}
w_2=-\textstyle\frac12\langle (x-\bar x)(p-\bar p)+(p-\bar p)(x-\bar
x)\rangle.
\end{equation}
\end{subequations}
Therefore,
\begin{equation}\label{83}
(\Delta x)^2=w_3+w_1, \quad (\Delta p)^2=w_3-w_1,
\end{equation}
and their time evolution is essentially governed by $E(t)$.

We will see that the pattern of motion does not have a simple
relation with the nature of the Hamiltonian. For a Hamiltonian with a
definite nature, say, elliptic, the wave packet may exhibit different
patterns of motion if some parameter in the Hamiltonian takes
different values. And, when the parameter goes through some critical
value, the motion changes from one pattern to another. This is
somewhat like phase transition, and seems not to be noticed before.

\subsection{$\bm n=\mathrm{constant}$}\label{s6.1}

The first case is the simplest one where $\bm n$ is a constant vector
and $\omega(t)$ is an arbitrary function of time. Let
\begin{equation}\label{84}
\varphi(t)=\int_0^t \omega(t')\; \rd t,
\end{equation}
and if ${\bm e}(t)={\bm e}[\varphi(t)]$ obeys
\begin{equation}\label{85}
{\bm e}'(\varphi)=-2{\bm n^g}\times {\bm e^g}(\varphi)
\end{equation}
where the prime indicates derivative with respect to $\varphi$, then
Eq. (\ref{17a}) is satisfied. This is similar to Eq. (\ref{5}) and
can be solved in the same way. The result depends on the nature of
$\bm n$, and will be given separately.

(1) If $\bm n^2=1$, the solution reads
\begin{equation}\label{86}
\bm e(t)=[\bm e_0 - (\bm e_0\cdot \bm n^g)\bm n] \cos2\varphi+ \bm
e_0^g \times\bm n^g \sin2\varphi+(\bm e_0\cdot \bm n^g)\bm n.
\end{equation}
To obtain the evolution matrix $E(t)$, we should write down the
components of $\bm n$:
\begin{equation}\label{87}
\bm n=(\sinh\xi_n\cos2\varphi_n,\sinh\xi_n\sin2\varphi_n,
\cosh\xi_n),
\end{equation}
where $\xi_n$ and $\varphi_n$ are arbitrary. The matrix $E(t)$ is
somewhat complicated, but it can be written as the product of several
simple matrices
\begin{equation}\label{88}
E(t)=R(\varphi_n)S(\xi_n,1)W^{(a)}_+(\varphi)
S^{-1}(\xi_n,1)R^{-1}(\varphi_n),
\end{equation}
where
\begin{equation}\label{89}
W^{(a)}_+(\varphi)=
\begin{pmatrix}
  \cos2\varphi & -\sin2\varphi & 0 \\
  \sin2\varphi & \cos2\varphi  & 0 \\
  0 & 0 & 1 \\
\end{pmatrix},
\end{equation}
and the matrices $R$ and $S$, which will be repeatedly used below,
are defined by
\begin{equation}\label{90}
R(\varphi_n)=
\begin{pmatrix}
  \cos2\varphi_n & -\sin2\varphi_n & 0 \\
  \sin2\varphi_n & \cos2\varphi_n  & 0 \\
  0 & 0 & 1 \\
\end{pmatrix}, \quad
S(\xi_n, \epsilon)=
\begin{pmatrix}
  \epsilon\cosh\xi_n & 0 & \sinh\xi_n \\
  0 & 1 & 0 \\
  \sinh\xi_n & 0 & \epsilon\cosh\xi_n \\
\end{pmatrix},
\end{equation}
where $\epsilon=\pm 1$. The eigenvalues of $E(t)$ can be easily found
to be $\{1, \e^{\ri 2\varphi}, \e^{-\ri 2\varphi}\}$. The eigenvector
corresponding to the eigenvalue 1 is $\bm\eta=\bm n$. This can be
easily seen from Eq. (\ref{86}), where $\bm e_0=\bm n$ leads to $\bm
e(t)=\bm n$ at any time $t$. Because $\bm n^2=1$, there exist
normalizable cyclic solutions at any time interval $[0,\tau]$.
However, most of these cyclic solutions are trivial ones. Actually,
the time evolution operator can be obtained by direct integration in
this case:
\begin{equation}\label{91}
U(t)=\exp(-\ri\hbar^{-1}\varphi\bm K\cdot \bm n^g).
\end{equation}
If $\psi(0)$ is an eigenstate of $\bm K\cdot \bm n^g$, then $\psi(t)$
is different from $\psi(0)$ only by a phase factor, thus it is cyclic
in any time interval but is trivial. Only when $\varphi(\tau)/\pi$
takes rational numbers that we have nontrivial cyclic solutions in
$[0,\tau]$.

The evolution matrix $E_q(t)$ can be similarly found to be
\begin{equation}\label{92}
E_q(t)=R_q(\varphi_n)W^{(a)}_{q+}(\varphi)R_q^{-1}(\varphi_n),
\end{equation}
where
\begin{equation}\label{93}
W^{(a)}_{q+}(\varphi)=\begin{pmatrix}
  \cos\varphi & \exp(\xi_n)\sin\varphi \\
  -\exp(-\xi_n)\sin\varphi & \cos\varphi \\
\end{pmatrix},
\end{equation}
and
\begin{equation}\label{94}
R_q(\varphi_n)=
\begin{pmatrix}
  \cos\varphi_n & \sin\varphi_n \\
  -\sin\varphi_n & \cos\varphi_n \\
  \end{pmatrix}.
\end{equation}
Obviously, when $\varphi(\tau)=-N\pi$, $E_q(\tau)=(-1)^N$ and
$E(\tau)=1$. This includes the two most interesting cases discussed
in Sec. \ref{s5}.

The time evolution operator (\ref{91}) can also be obtained by our
method. Taking the simplest solution $\bm e(t)=\bm n$, we have
$\xi(t)=\xi_0=\xi_n$, $\phi(t)=\phi_0=2\varphi_n$, and
$\alpha(t)=-\varphi(t)$. When these are substituted into Eq.
(\ref{34b}), we obtain the above result. Moreover, the above
condition $\varphi(\tau)=-N\pi$ for $E_q(\tau)=(-1)^N$ is equivalent
to $\alpha(\tau)=N\pi$, as expected.

To verify Eq. (\ref{67}) by explicit calculations, we take $\bm
n=(0,0,1)$, then $U(t)=\exp[-\ri\hbar^{-1}\varphi(t) K_3]$. When
$\varphi(\tau)=-2N\pi$, we have $U(\tau)=\e^{\ri N\pi}$. In the time
interval $[0,\tau]$, all solutions are cyclic with $\delta=N\pi$, mod
$2\pi$. Let the initial state have $\bm u(0)=\hbar u_0 \bm e_0$,
where $\bm e_0=(\sinh\xi_0\cos\phi_0, \sinh\xi_0\sin\phi_0,
\cosh\xi_0)$. Then $\bm u(t)=\hbar u_0 \bm e(t)$, where $\bm e(t)=
\bm(\sinh\xi_0\cos(2\varphi+\phi_0), \sinh\xi_0\sin(2\varphi+\phi_0),
\cosh\xi_0\bm)$. It is then easy to find that $\beta=2N\pi u_0
\cosh\xi_0$, and $\gamma=N\pi(1-2u_0 \cosh\xi_0)$. On the other hand,
the Hannay angle can be found to be $\Delta\theta_{\mathrm g}=2N\pi
(\cosh\xi_0-1)$. With these results it is easy to see that Eq.
(\ref{67}) is valid.

(2) If $\bm n^2=-1$, the solution reads
\begin{equation}\label{95}
\bm e(t)=[\bm e_0 + (\bm e_0\cdot \bm n^g)\bm n] \cosh2\varphi+ \bm
e_0^g \times\bm n^g \sinh2\varphi-(\bm e_0\cdot \bm n^g)\bm n.
\end{equation}
To obtain the evolution matrix $E(t)$, we should write down the
components of $\bm n$:
\begin{equation}\label{96}
\bm n=(\cosh\xi_n\cos2\varphi_n,\cosh\xi_n\sin2\varphi_n,
\sinh\xi_n),
\end{equation}
where $\xi_n$ and $\varphi_n$ are arbitrary. The matrix $E(t)$ is
given by
\begin{equation}\label{97}
E(t)=R(\varphi_n)S(\xi_n,1)W^{(a)}_-(\varphi)S^{-1}(\xi_n,1)R^{-1}(\varphi_n),
\end{equation}
where
\begin{equation}\label{98}
W^{(a)}_-(\varphi)=
\begin{pmatrix}
1 & 0 & 0 \\
0 & \cosh2\varphi & -\sinh2\varphi \\
0 & -\sinh2\varphi & \cosh2\varphi \\
\end{pmatrix}.
\end{equation}
The eigenvalues of $E(t)$ can be easily found to be $\{1, \e^{2
\varphi}, \e^{-2\varphi}\}$. The eigenvector corresponding to the
eigenvalue 1 is $\bm\eta=\bm n$. Because $\bm n^2=-1$, there exists
no normalizable cyclic solution at any time interval $[0,\tau]$. The
evolution matrix $E_q(t)$ can be found to be
\begin{equation}\label{99}
E_q(t)=R_q(\varphi_n)W^{(a)}_{q-}(\varphi)R_q^{-1}(\varphi_n),
\end{equation}
where
\begin{equation}\label{100}
W^{(a)}_{q-}(\varphi)=\begin{pmatrix}
  \cosh\varphi & \exp(\xi_n)\sinh\varphi \\
  \exp(-\xi_n)\sinh\varphi & \cosh\varphi \\
\end{pmatrix}.
\end{equation}
In this case $E_q(\tau)$ cannot take the value $\pm1$, otherwise one
should obtain $W^{(a)}_{q-}(\varphi)=\pm1$. The latter is obviously
impossible, unless $\varphi(\tau)$ returns to 0 but this is not of
interest.

(3) If $\bm n^2=0$, the solution reads
\begin{equation}\label{101}
\bm e(t)=2\varphi^2 (\bm e_0\cdot \bm n^g)\bm n + 2\varphi\bm e_0^g
\times\bm n^g +\bm e_0.
\end{equation}
Up to a constant factor, which may be absorbed in $\omega(t)$, the
components of $\bm n$ may be written as:
\begin{equation}\label{102}
\bm n=(\cos2\varphi_n,\sin2\varphi_n, 1),
\end{equation}
where $\varphi_n$ is arbitrary. The matrix $E(t)$ is given by
\begin{equation}\label{103}
E(t)=R(\varphi_n)W^{(a)}_0(\varphi)R^{-1}(\varphi_n),
\end{equation}
where
\begin{equation}\label{104}
W^{(a)}_0(\varphi)=
\begin{pmatrix}
1-2\varphi^2 & -2\varphi & 2\varphi^2 \\
2\varphi & 1 & -2\varphi \\
-2\varphi^2 & -2\varphi & 1+2\varphi^2 \\
\end{pmatrix}.
\end{equation}
The eigenvalues of $E(t)$ can be easily found to be $\{1, 1, 1\}$,
but there is only one eigenvector $\bm\eta=\bm n$ if $\varphi\ne0$.
Thus we find an example where all eigenvalues are 1 but $E(t)$ cannot
be diagonalized, let alone be a unit matrix. Because $\bm n^2=0$,
there exists no normalizable cyclic solution at any time interval
$[0,\tau]$. The evolution matrix $E_q(t)$ can be found to be
\begin{equation}\label{105}
E_q(t)=R_q(\varphi_n)W^{(a)}_{q0}(\varphi)R_q^{-1}(\varphi_n),
\end{equation}
where
\begin{equation}\label{106}
W^{(a)}_{q0}(\varphi)=\begin{pmatrix}
  1 & 2\varphi \\
  0 & 1 \\
\end{pmatrix}.
\end{equation}
In this case $E_q(\tau)$ cannot take the value $\pm1$ either, unless
$\varphi(\tau)$ returns to 0 but this is not of interest.

From the above results we see that the motion is finite when the
Hamiltonian is elliptic and infinite (assume that $\varphi(t)$ can
reach arbitrarily large values, for example, $\varphi(t)$ is
proportional to $t$) in the other cases. But note that $E(t)$ and
$E_q(t)$ are polynomials of $\varphi$ in the critical case and are
exponential functions of it in the hyperbolic case. If $\bm n$ is
time dependent, however, the situation is not so simple. We will see
that for a Hamiltonian of a definite nature, all patterns of motion
are possible.

\subsection{$\bm n=(n_1\cos2\varphi, n_1\sin2\varphi,
n_3)$}\label{s6.2}

Here $n_1$ and $n_3$ are constants, satisfying $\bm
n^2=n_3^2-n_1^2=\pm1$ or 0, and $\varphi=\varphi(t)$ is an arbitrary
function of $t$ with $\varphi(0)=0$. For convenience, we assume
henceforward that $\varphi(t)$ increases with $t$ monotonically and
goes to infinity when $t$ does. If $\omega(t)=\lambda \dot
\varphi(t)$ where $\lambda$ is a constant, analytic solutions are
available.

By making a time-dependent unitary transformation $U(t)=\exp [-\ri
\hbar^{-1}\varphi K_3]\tilde U(t)$, it is not difficult to obtain
\begin{equation}\label{107}
U(t)=\exp (-\ri \hbar^{-1}\varphi K_3) \exp (-\ri \hbar^{-1} \Lambda
\varphi \bm K \cdot \bm n_f^g),
\end{equation}
where $\Lambda>0$ and $\bm n_f^2=\pm 1$ or 0, and
\begin{equation}\label{108}
\Lambda\bm n_f=(\lambda n_1, 0, \lambda n_3 -1).
\end{equation}
If $\bm n_f^2=1$, we see that cyclic solutions are available in
$[0,\tau]$ if the initial state is an eigenstate of $\bm K \cdot \bm
n_f^g$ and $\varphi(\tau)=N\pi$, and more general cyclic solutions
are available in some time interval if $\Lambda$ happens to be a
rational number. However, for other time interval, or when $\bm
n_f^2=-1$ or 0, the above time evolution operator, though explicit,
does not tell much about cyclic solutions. Consequently, it is
necessary to study the matrix $E(t)$ and $E_q(t)$.

If ${\bm e}(t)={\bm e}[\varphi(t)]$ obeys
\begin{equation}\label{109}
{\bm e}'(\varphi)=-2\lambda {\bm n^g}(\varphi)\times {\bm
e^g}(\varphi)
\end{equation}
then Eq. (\ref{17a}) is satisfied. Now we make a time-dependent
linear transformation, written in column vector form:
\begin{equation}\label{110}
e(\varphi)=R(\varphi)f(\varphi),
\end{equation}
where the matrix $R$ is defined in Eq. (\ref{90}), but now the
independent variable $\varphi$ is time dependent. The reduced
equation for $f(\varphi)$, written in ordinary vector form, reads
\begin{equation}\label{111}
{\bm f}'(\varphi)=-2\Lambda {\bm n_f^g}\times {\bm f^g}(\varphi).
\end{equation}
Now $\bm n_f$ is a constant vector, so the reduced equation is easy
to solve. The form of the solution depends on the nature of $\bm
n_f$, and several cases should be treated separately.

(1) If $\bm n^2=1$ and $\lambda\in (-\infty, n_3-|n_1|)
\cup(n_3+|n_1|,+\infty)$, or $\bm n^2=-1$ and $\lambda\in
(-n_3-|n_1|, -n_3+|n_1|)$, or $\bm n^2=0$ and $\lambda n_3<1/2$, then
$(\lambda n_3-1)^2>(\lambda n_1)^2$. In this case we define
\begin{equation}\label{112}
\Lambda=\sqrt{(\lambda n_3-1)^2-(\lambda n_1)^2}, \quad \cosh\xi_n
=|\lambda n_3-1|/\Lambda, \quad \sinh\xi_n=\lambda n_1/\Lambda,
\end{equation}
then
\begin{equation}\label{113}
\bm n_f=(\sinh\xi_n,0, \epsilon\cosh\xi_n),
\end{equation}
where $\epsilon=\epsilon(\lambda n_3-1)$ is the sign function of
$\lambda n_3-1$, that is, it equals 1 ($-1$) if $\lambda n_3-1>0$
($<0$). The solution is
\begin{equation}\label{114}
E(t)=R(\varphi)S(\xi_n,\epsilon)W^{(b)}_{+}(\varphi)S^{-1}(\xi_n,\epsilon),
\quad E_q(t)=R_q(\varphi) W^{(b)}_{q+}(\varphi),
\end{equation}
where
\begin{equation}\label{115}
W^{(b)}_+(\varphi)=
\begin{pmatrix}
  \cos2\Lambda\varphi & -\sin2\Lambda\varphi & 0 \\
  \sin2\Lambda\varphi & \cos2\Lambda\varphi  & 0 \\
  0 & 0 & 1 \\
\end{pmatrix}, \quad
W^{(b)}_{q+}(\varphi)=\begin{pmatrix}
\cos\Lambda\varphi&\epsilon\exp(\epsilon\xi_n)\sin\Lambda\varphi\\
-\epsilon\exp(-\epsilon\xi_n)\sin\Lambda\varphi&\cos\Lambda\varphi\\
\end{pmatrix},
\end{equation}
and the matrix $R_q$ is defined in Eq. (\ref{94}), but now the
independent variable $\varphi$ is time dependent. In this case the
motion is finite, no matter the Hamiltonian is elliptic, hyperbolic
or critical, as long as $\lambda$ belongs to the appropriate region.

The eigenvector of $E(t)$ corresponding to the eigenvalue 1 is (not
normalized)
\begin{equation}\label{116}
\bm\eta(t)=(\epsilon\sinh\xi_n \sin\Lambda\varphi\cos\varphi,
\epsilon\sinh\xi_n \sin\Lambda\varphi\sin\varphi, \cosh\xi_n
\sin\Lambda\varphi\cos\varphi+\epsilon
\cos\Lambda\varphi\sin\varphi).
\end{equation}
If $\sin\varphi=0$ but $\sin\Lambda\phi\ne 0$ at the time $\tau$, or
$\sin\Lambda\varphi=0$ but $\sin\phi\ne 0$ (if both are zero, then
$E(\tau)=1$), then $\bm\eta^2(\tau)>0$ and there are cyclic solutions
in the time interval $[0,\tau]$. In the general case, we have
\begin{equation}\label{117}
\bm\eta^2(t)=(\sinh^2\xi_n\sin^2\Lambda\varphi+1)\sin^2[\varphi+
\epsilon \arctan(\cosh\xi_n \tan\Lambda\varphi)]
-\sinh^2\xi_n\sin^2\Lambda\varphi.
\end{equation}
This may be either positive or negative. If at the time $\tau$ the
quantity in the square bracket is close to $(N+1/2)\pi$, then
$\bm\eta^2(\tau)>0$, and there are cyclic solutions in the time
interval $[0,\tau]$. If the quantity is close to $N\pi$, and
$\sin\Lambda\varphi$ is not too small, then $\bm\eta^2(\tau)<0$, and
there is no cyclic solution in the time interval $[0,\tau]$.

If $\Lambda$ is a rational number, then there exist some $\tau$ such
that $E_q(\tau)=\pm 1$ and $E(\tau)=1$. This is the case where more
general cyclic solutions are available.

The time evolution operator can be easily obtained by our formalism
in this case. We take the simple solution $\bm
e(t)=(\epsilon\sinh\xi_n\cos2\varphi, \epsilon\sinh\xi_n\sin2\varphi,
\cosh\xi_n)$ and using Eq. (\ref{34a}), after some operator algebras,
we arrive at the result (\ref{107}). For the subsequent cases,
however, our formalism is not convenient in calculating the time
evolution operator, since it is not easy to find a solution that is
sufficiently simple and satisfies Eq. (\ref{17c}). Instead, the
time-dependent unitary transformation that leads to the result
(\ref{107}) is convenient in this respect. However, as pointed out
before, the result (\ref{107}) does not tell much about cyclic
solutions.

(2) If $\bm n^2=1$ and $\lambda\in (n_3-|n_1|, n_3+|n_1|)$, or $\bm
n^2=-1$ and $\lambda\in (-\infty, -n_3-|n_1|)
\cup(-n_3+|n_1|,+\infty)$, or $\bm n^2=0$ and $\lambda n_3>1/2$, then
$(\lambda n_3-1)^2<(\lambda n_1)^2$. In this case we define
\begin{equation}\label{118}
\Lambda=\sqrt{(\lambda n_1)^2-(\lambda n_3-1)^2}, \quad \cosh\xi_n
=|\lambda n_1|/\Lambda, \quad \sinh\xi_n=(\lambda n_1-1)/\Lambda,
\end{equation}
then
\begin{equation}\label{119}
\bm n_f=(\epsilon\cosh\xi_n,0,\sinh\xi_n),
\end{equation}
where $\epsilon=\epsilon(\lambda n_1)$. The solution is
\begin{equation}\label{120}
E(t)=R(\varphi)S(\xi_n,\epsilon)W^{(b)}_{-}(\varphi)S^{-1}(\xi_n,\epsilon),
\quad E_q(t)=R_q(\varphi) W^{(b)}_{q-}(\varphi),
\end{equation}
where
\begin{equation}\label{121}
W^{(b)}_-(\varphi)=
\begin{pmatrix}
1 & 0 & 0\\
0 & \cosh2\Lambda\varphi & -\sinh2\Lambda\varphi \\
0 & -\sinh2\Lambda\varphi & \cosh2\Lambda\varphi \\
\end{pmatrix}, \quad
W^{(b)}_{q-}(\varphi)=\begin{pmatrix}
  \cosh\Lambda\varphi & \epsilon\exp(\epsilon\xi_n)\sinh\Lambda\varphi \\
  \epsilon\exp(-\epsilon\xi_n)\sinh\Lambda\varphi & \cosh\Lambda\varphi \\
\end{pmatrix}.
\end{equation}
In this case the motion is oscillating with exponentially increasing
amplitude, no matter the Hamiltonian is elliptic, hyperbolic or
critical, as long as $\lambda$ belongs to the appropriate region.

The eigenvector of $E(t)$ corresponding to the eigenvalue 1 is (not
normalized)
\begin{equation}\label{122}
\bm\eta(t)=(\epsilon\cosh\xi_n \sinh\Lambda\varphi\cos\varphi,
\epsilon\cosh\xi_n \sinh\Lambda\varphi\sin\varphi, \sinh\xi_n
\sinh\Lambda\varphi\cos\varphi+ \cosh\Lambda\varphi\sin\varphi).
\end{equation}
If $\sin\varphi=0$ but $\varphi\ne 0$ at the time $\tau$, then
$\bm\eta^2(\tau)<0$ and there is no cyclic solution in the time
interval $[0,\tau]$. In the general case, we have
\begin{equation}\label{123}
\bm\eta^2(t)=(\cosh^2\xi_n\sinh^2\Lambda\varphi+1)\sin^2[\varphi+
\arctan(\sinh\xi_n \tanh\Lambda\varphi)]
-\cosh^2\xi_n\sinh^2\Lambda\varphi.
\end{equation}
This may be either positive or negative. If at the time $\tau$ the
quantity in the square bracket is close to $(N+1/2)\pi$, then
$\bm\eta^2(\tau)>0$, and there are cyclic solutions in the time
interval $[0,\tau]$. If the quantity is close to $N\pi$, and
$\varphi$ is not too small, then $\bm\eta^2(\tau)<0$, and there is no
cyclic solution in the time interval $[0,\tau]$. In the present case
there exists no $\tau$ such that $\varphi(\tau)\ne 0$ and
$E_q(\tau)=\pm 1$.

(3) If $\bm n^2=1$ and $\lambda=n_3\pm |n_1|$, or $\bm n^2=-1$ and
$\lambda=-n_3\pm |n_1|$, or $\bm n^2=0$ and $\lambda n_3=1/2$, then
$(\lambda n_3-1)^2=(\lambda n_1)^2$. Let $\epsilon=(\lambda
n_3-1)/\lambda n_1=\pm 1$. The solution is
\begin{equation}\label{124}
E(t)=R(\varphi)D^{(b)}(\varphi), \quad E_q(t)=R_q(\varphi)
W^{(b)}_{q0}(\varphi),
\end{equation}
where
\begin{equation}\label{125}
D^{(b)}(\varphi)=
\begin{pmatrix}
1-2(\lambda n_1)^2\varphi^2 & -2\epsilon\lambda n_1\varphi &%
2\epsilon(\lambda n_1)^2\varphi^2\\
2\epsilon\lambda n_1\varphi & 1 & -2\lambda n_1\varphi \\
-2\epsilon(\lambda n_1)^2\varphi^2 & -2\lambda n_1\varphi &%
1+2(\lambda n_1)^2\varphi^2 \\
\end{pmatrix}, \quad
W^{(b)}_{q0}(\varphi)=\begin{pmatrix}
  1 & (1+\epsilon)\lambda n_1\varphi \\
  (1-\epsilon)\lambda n_1\varphi & 1 \\
\end{pmatrix}.
\end{equation}
In this case the motion is oscillating with polynomially increasing
amplitude, no matter the Hamiltonian is elliptic, hyperbolic or
critical, as long as $\lambda$ takes the appropriate value.

The eigenvector of $E(t)$ corresponding to the eigenvalue 1 is (not
normalized)
\begin{equation}\label{126}
\bm\eta(t)=(\lambda n_1\varphi\cos\varphi, \lambda
n_1\varphi\sin\varphi, \epsilon\lambda n_1\varphi\cos\varphi+
\sin\varphi).
\end{equation}
If $\sin\varphi=0$ at the time $\tau$, then $\bm\eta^2(\tau)=0$ and
there is no cyclic solution in the time interval $[0,\tau]$. In the
general case, we have
\begin{equation}\label{127}
\bm\eta^2(t)=(\lambda^2 n_1^2\varphi^2+1)\sin^2[\varphi+
\epsilon\arctan(\lambda n_1\varphi)] -\lambda^2 n_1^2\varphi^2.
\end{equation}
This may be either positive or negative. If at the time $\tau$ the
quantity in the square bracket is close to $(N+1/2)\pi$, then
$\bm\eta^2(\tau)>0$, and there are cyclic solutions in the time
interval $[0,\tau]$. If the quantity is close to $N\pi$, and
$\varphi$ is not too small, then $\bm\eta^2(\tau)<0$, and there is no
cyclic solution in the time interval $[0,\tau]$. In the present case
there exists no $\tau$ such that $\varphi(\tau)\ne 0$ and
$E_q(\tau)=\pm 1$.

It should be noted that all the three eigenvalues of $E(t)$ are 1
when $\varphi(t)=N\pi$, and all the two eigenvalues of $E_q(t)$ are 1
when $\varphi(t)=2N\pi$, but they are not unit matrices unless $N=0$.

From the above results we see that for a Hamiltonian of a definite
nature, say, elliptic, the motion may have all possible patterns,
depending on the value of $\lambda$. When $\lambda< n_3-|n_1|$ or
$\lambda> n_3+|n_1|$ the motion is finite. When $n_3-|n_1|<\lambda<
n_3+|n_1|$, the motion is oscillating with exponentially increasing
amplitude. At the two critical values $\lambda=n_3\pm |n_1|$ the
motion is oscillating with polynomially increasing amplitude.
Something like phase transition happens here. When $\lambda$ goes
through the critical values, the motion changes from one pattern to
another, and at the critical values the motion exhibits an
independent pattern. Similar situation will be seen in the next
subsection.

A common feature of the three different cases above is that in any
time interval $(\tau_0,+\infty)$ with $\tau_0>0$ one can always find
some $\tau$ such that cyclic solutions are available in $[0,\tau]$.

\subsection{$\bm n=(n_1, n_3\sinh2\varphi,
n_3\cosh2\varphi)$}\label{s6.3}

As before, $n_1$ and $n_3$ are constants, satisfying $\bm
n^2=n_3^2-n_1^2=\pm1$ or 0, and $\varphi=\varphi(t)$ has the same
property as in subsection \ref{s6.2}. If $\omega(t)=\lambda \dot
\varphi(t)$ where $\lambda$ is a constant, analytic solutions are
available. The time evolution operator can be obtained in a way
similar to that in subsection \ref{s6.2}. The result is
\begin{equation}\label{128}
U(t)=\exp (-\ri \hbar^{-1}\varphi K_1) \exp (-\ri \hbar^{-1} \Lambda
\varphi \bm K \cdot \bm n_f^g),
\end{equation}
where $\Lambda>0$ and $\bm n_f^2=\pm 1$ or 0, and
\begin{equation}\label{129}
\Lambda\bm n_f=(\lambda n_1+1, 0, \lambda n_3).
\end{equation}
Compared with the case in subsection \ref{s6.2}, here still less
about cyclic solutions can be seen from the above result.
Consequently, it is necessary to study the matrix $E(t)$ and
$E_q(t)$.

If ${\bm e}(t)={\bm e}[\varphi(t)]$ obeys an equation of the form
(\ref{109}), then Eq. (\ref{17a}) is satisfied. As before we make a
time-dependent linear transformation, written in column vector form:
\begin{equation}\label{130}
e(\varphi)=T(\varphi)f(\varphi),
\end{equation}
where the matrix $T$, and another matrix $T_q$ that will be used
below, are defined as
\begin{equation}\label{131}
T(\varphi)=
\begin{pmatrix}
  1 & 0 & 0 \\
  0 & \cosh2\varphi & \sinh2\varphi \\
  0 & \sinh2\varphi & \cosh2\varphi \\
\end{pmatrix}, \quad
T_q(\varphi)=
\begin{pmatrix}
  \cosh\varphi & -\sinh\varphi \\
  -\sinh\varphi & \cosh\varphi \\
\end{pmatrix}.
\end{equation}
The reduced equation for $f(\varphi)$, written in ordinary vector
form, has the form of Eq. (\ref{111}), but where $\Lambda {\bm n_f}$
is given by Eq. (\ref{129}). As before, several cases are to be
treated separately.

(1) If $\bm n^2=1$ and $\lambda\in (-\infty, n_1-|n_3|)
\cup(n_1+|n_3|,+\infty)$, or $\bm n^2=-1$ and $\lambda\in
(-n_1-|n_3|, -n_1+|n_3|)$, or $\bm n^2=0$ and $\lambda n_1<-1/2$,
then $(\lambda n_3)^2>(\lambda n_1+1)^2$. In this case we define
\begin{equation}\label{132}
\Lambda=\sqrt{(\lambda n_3)^2-(\lambda n_1+1)^2}, \quad \cosh\xi_n
=|\lambda n_3|/\Lambda, \quad \sinh\xi_n=(\lambda n_1+1)/\Lambda,
\end{equation}
then $\bm n_f$ has the form in Eq. ({113}), but
$\epsilon=\epsilon(\lambda n_3)$. The solution is
\begin{equation}\label{133}
E(t)=T(\varphi)S(\xi_n,\epsilon)W^{(b)}_{+}(\varphi)S^{-1}(\xi_n,\epsilon),
\quad E_q(t)=T_q(\varphi) W^{(b)}_{q+}(\varphi),
\end{equation}
where $W^{(b)}_+(\varphi)$ and $W^{(b)}_{q+}(\varphi)$ have been
given in Eq. ({115}). In this case the motion is oscillating with
exponentially increasing amplitude.

The eigenvector of $E(t)$ corresponding to the eigenvalue 1 is (not
normalized)
\begin{equation}\label{134}
\bm\eta(t)=\bm(\epsilon(\sinh\xi_n\cosh\varphi \sin\Lambda\varphi
-\sinh\varphi\cos\Lambda\varphi), \cosh\xi_n\sinh\varphi
\sin\Lambda\varphi, \cosh\xi_n \cosh\varphi \sin\Lambda\varphi\bm).
\end{equation}
If $\sin\Lambda\varphi=0$ but $\varphi\ne 0$ at the time $\tau$, then
$\bm\eta^2(\tau)<0$ and there are no cyclic solution in the time
interval $[0,\tau]$. In the general case, we have
\begin{equation}\label{135}
\bm\eta^2(t)=\cosh^2\xi_n\sin^2\Lambda\varphi-
(\cosh^2\xi_n\cosh^2\varphi-1)\sin^2[\Lambda\varphi-\arctan(
\tanh\varphi/\sinh\xi_n)].
\end{equation}
This may be either positive or negative. For large $\varphi$ the
second term is large as long as the quantity in the square bracket is
not close to $N\pi$, then $\bm\eta^2(\tau)<0$, and there is no cyclic
solution. However, if $\varphi(\tau)$ is close to $\varphi_0$ where
$\varphi_0$ is the root of
$\tan\Lambda\varphi=\tanh\varphi/\sinh\xi_n$, the second term is very
small and $\bm\eta^2(\tau)>0$, then we have cyclic solutions in the
time interval $[0,\tau]$. The above transcendental equation has
infinitely many roots, therefore in any time interval
$(\tau_0,+\infty)$ with $\tau_0>0$ we can always find some $\tau$
such that cyclic solutions are available in $[0,\tau]$. In the
present case there exists no $\tau$ such that $\varphi(\tau)\ne 0$
and $E_q(\tau)=\pm 1$.

(2) If $\bm n^2=1$ and $\lambda\in (n_1-|n_3|, n_1+|n_3|)$, or $\bm
n^2=-1$ and $\lambda\in (-\infty,-n_1-|n_3|)\cup(-n_1+|n_3|,\infty)$,
or $\bm n^2=0$ and $\lambda n_1>-1/2$, then $(\lambda n_3)^2<(\lambda
n_1+1)^2$. In this case we define
\begin{equation}\label{136}
\Lambda=\sqrt{(\lambda n_1+1)^2-(\lambda n_3)^2}, \quad \cosh\xi_n
=|\lambda n_1+1|/\Lambda, \quad \sinh\xi_n=\lambda n_3/\Lambda,
\end{equation}
then $\bm n_f$ has the form in Eq. ({119}), but
$\epsilon=\epsilon(\lambda n_1+1)$. The solution is
\begin{equation}\label{137}
E(t)=T(\varphi)S(\xi_n,\epsilon)W^{(b)}_{-}(\varphi)S^{-1}(\xi_n,\epsilon),
\quad E_q(t)=T_q(\varphi) W^{(b)}_{q-}(\varphi),
\end{equation}
where $W^{(b)}_-(\varphi)$ and $W^{(b)}_{q-}(\varphi)$ have been
given in Eq. ({121}). In this case the motion is exponentially
infinite.

The eigenvector of $E(t)$ corresponding to the eigenvalue 1 is (not
normalized)
\begin{equation}\label{138}
\bm\eta(t)=(\epsilon\cosh\xi_n\cosh\varphi \sinh\Lambda\varphi
-\sinh\varphi\cosh\Lambda\varphi, \sinh\xi_n\sinh\varphi
\sinh\Lambda\varphi, \sinh\xi_n \cosh\varphi \sinh\Lambda\varphi),
\end{equation}
which gives
\begin{equation}\label{139}
\bm\eta^2(t)=\sinh^2\xi_n\sinh^2\Lambda\varphi-
(\cosh\xi_n\cosh\varphi \sinh\Lambda\varphi
-\epsilon\sinh\varphi\cosh\Lambda\varphi)^2.
\end{equation}
If $\epsilon=-1$, it is obvious that the second term is larger and
$\bm\eta^2(\tau)<0$, and thus there is no cyclic solution in any time
interval $[0,\tau]$. If $\epsilon=1$, the above result can be recast
in the form
\begin{equation}\label{140}
\bm\eta^2(t)=\sinh^2\xi_n\sinh^2\Lambda\varphi-
(\sinh^2\xi_n\cosh^2\varphi+1) \sinh^2[\Lambda\varphi
-\mathrm{Artanh}(\tanh\varphi/\cosh\xi_n)].
\end{equation}
In this case, cyclic solutions are possible. For example, when
$\Lambda<1$, the transcendental equation $\tanh\varphi=\cosh\xi_n
\tanh\Lambda\varphi$ has one root $\varphi_0$. When $\varphi(\tau)$
is close to $\varphi_0$, the second term is very small and
$\bm\eta^2(\tau)>0$. Then we can have cyclic solutions in $[0,\tau]$.
Another possible case is $\varphi(\tau)\ll 1$, which also leads to
$\bm\eta^2(\tau)>0$ if $\exp(|\xi_n|)>\Lambda$. Then we also have
cyclic solutions in $[0,\tau]$. However, when $\varphi(\tau)$ is
large, it is easy to see from Eq. (\ref{139}) that
$\bm\eta^2(\tau)<0$ regardless of the parameters $\xi_n$ and
$\Lambda$. Then there is no cyclic solution in $[0,\tau]$. In other
words, no initial state can return to itself at a large time. This is
rather different from the last case. As before, there exists no
$\tau$ such that $\varphi(\tau)\ne 0$ and $E_q(\tau)=\pm 1$.

(3) If $\bm n^2=1$ and $\lambda=n_1\pm |n_3|$, or $\bm n^2=-1$ and
$\lambda=-n_1\pm |n_3|$, or $\bm n^2=0$ and $\lambda n_1=-1/2$, then
$(\lambda n_3)^2=(\lambda n_1+1)^2$. Let $\epsilon=(\lambda
n_1+1)/\lambda n_3=\pm1$. The solution is
\begin{equation}\label{141}
E(t)=T(\varphi)D^{(c)}(\varphi), \quad E_q(t)=T_q(\varphi)
W^{(c)}_{q0}(\varphi),
\end{equation}
where $D^{(c)}(\varphi)$ and $W^{(c)}_{q0}(\varphi)$ can be obtained
from $D^{(b)}(\varphi)$ and $W^{(b)}_{q0}(\varphi)$ in Eq.
(\ref{125}), respectively, by substituting $\epsilon\lambda n_3$ for
$\lambda n_1$. In this case the motion is also exponentially
infinite, but different from the last case. Indeed, both
$W^{(b)}_-(\varphi)$ and $W^{(b)}_{q-}(\varphi)$ in Eq. (\ref{137})
involve exponential functions of $\varphi$, while $D^{(c)}(\varphi)$
and $W^{(c)}_{q0}(\varphi)$ involve polynomials of it.

The eigenvector of $E(t)$ corresponding to the eigenvalue 1 is (not
normalized)
\begin{equation}\label{142}
\bm\eta(t)=(\epsilon\lambda n_3\varphi\cosh\varphi-\sinh\varphi,
\lambda n_3\varphi\sinh\varphi, \lambda n_3\varphi\cosh\varphi),
\end{equation}
which gives
\begin{equation}\label{143}
\bm\eta^2(t)=(\lambda n_3\varphi)^2-(\epsilon\lambda
n_3\varphi\cosh\varphi-\sinh\varphi)^2.
\end{equation}
If $\epsilon\lambda n_3<0$, it is obvious that the second term is
larger and $\bm\eta^2(\tau)<0$, and thus there is no cyclic solution
in any time interval $[0,\tau]$. If $\epsilon\lambda n_3>0$, cyclic
solutions are possible. For example, when $|\lambda n_3|<1$, the
transcendental equation $\tanh\varphi=\epsilon\lambda n_3\varphi$ has
one nonzero root $\varphi_0$. When $\varphi(\tau)$ is close to
$\varphi_0$, the second term is very small and $\bm\eta^2(\tau)>0$.
Then we can have cyclic solutions in $[0,\tau]$. Another possible
case is $\varphi(\tau)\ll 1$, which also leads to $\bm\eta^2(\tau)>0$
and we have cyclic solutions in $[0,\tau]$. However, when
$\varphi(\tau)$ is large, it is easy to see that $\bm\eta^2(\tau)<0$
regardless of the parameters $n_3$ and $\lambda$. Then there is no
cyclic solution in $[0,\tau]$. In other words, no initial state can
return to itself at a large time. This is similar to the last case.
As before, there exists no $\tau$ such that $\varphi(\tau)\ne 0$ and
$E_q(\tau)=\pm 1$.

In this subsection we also see that different values of $\lambda$
correspond to different patterns of motion, for a Hamiltonian of a
definite nature. But the cases are rather different from those in
subsection \ref{s6.2}. In some of the cases here, there is no cyclic
solution in any time interval.

\subsection{$\bm n=(n_1, n_3\cosh2\varphi,
n_3\sinh2\varphi)$}\label{s6.4}

Here $n_1$ and $n_3$ are constants, satisfying $\bm
n^2=-n_3^2-n_1^2=-1$. This is different from the previous cases,
since $\bm n^2$ cannot take 1 and 0. $\varphi=\varphi(t)$ has the
same property as before. If $\omega(t)=\lambda \dot \varphi(t)$ where
$\lambda$ is a constant, analytic solutions are available. The time
evolution operator is given by Eq. (\ref{128}) but now
\begin{equation}\label{144}
\Lambda\bm n_f=(\lambda n_1+1, \lambda n_3, 0).
\end{equation}
The equation for ${\bm e}(t)$ can be solved in a way similar to that
in subsection \ref{s6.3}. We define
\begin{equation}\label{145}
\Lambda=\sqrt{(\lambda n_1+1)^2+(\lambda n_3)^2}, \quad \cos2
\varphi_n=(\lambda n_1+1)/\Lambda, \quad \sin2\varphi_n= \lambda
n_3/\Lambda,
\end{equation}
then $\bm n_f=(\cos2 \varphi_n, \sin2\varphi_n, 0)$. The solution is
\begin{equation}\label{146}
E(t)=T(\varphi)R(\varphi_n)W^{(b)}_{-}(\varphi)R^{-1}(\varphi_n),
\quad E_q(t)=T_q(\varphi) R_q(\varphi_n) W^{(d)}_{q}(\varphi)
R_q^{-1}(\varphi_n),
\end{equation}
where $W^{(b)}_{-}(\varphi)$ is given in Eq. (\ref{121}) and
\begin{equation}\label{147}
W^{(d)}_{q}(\varphi)=\begin{pmatrix}
  \cosh\Lambda\varphi & \sinh\Lambda\varphi \\
  \sinh\Lambda\varphi & \cosh\Lambda\varphi \\
\end{pmatrix}.
\end{equation}
Obviously, the motion is exponentially infinite.

The eigenvector of $E(t)$ corresponding to the eigenvalue 1 is (not
normalized)
\begin{equation}\label{148}
\bm\eta(t)=(\cos2\varphi_n\cosh\varphi \sinh\Lambda\varphi
-\sinh\varphi\cosh\Lambda\varphi, \sin2\varphi_n\cosh\varphi
\sinh\Lambda\varphi, \sin2\varphi_n \sinh\varphi
\sinh\Lambda\varphi),
\end{equation}
which gives
\begin{equation}\label{149}
\bm\eta^2(t)=-\sin^2 2\varphi_n\sinh^2\Lambda\varphi-
(\cos2\varphi_n\cosh\varphi \sinh\Lambda\varphi
-\sinh\varphi\cosh\Lambda\varphi)^2.
\end{equation}
It is obvious that $\bm\eta^2(\tau)<0$ if $\varphi(\tau)\ne 0$, and
there is no cyclic solution in any time interval $[0,\tau]$. In other
words, in this case no initial state can return to itself at a later
time. As before, there exists no $\tau$ such that $\varphi(\tau)\ne
0$ and $E_q(\tau)=\pm 1$.

\section{Summary}\label{s7}

In this paper we develop a method for solving the Schr\"odinger
equation of the generalized time-dependent harmonic oscillator. This
method, though not always convenient in practical calculation of the
time evolution operator, is very suitable for the study of cyclic
solutions and geometric phases. We concentrate our attention on
Hamiltonian of general time dependence and cyclic solutions in the
time interval $[0,\tau]$ with an arbitrarily given $\tau$. A
necessary and sufficient condition for the existence of cyclic
solutions in such time intervals is given. There may exist some time
interval in which more solutions are cyclic. This includes several
cases among which two are of more interest. In one of these cases all
solutions are cyclic, and in another all solutions with definite
parity are cyclic. Criterion for the appearance of such cases are
given. The proportional relation between the nonadiabatic geometric
phase and the classical Hannay angle is reestablished. However, this
holds only for cyclic solutions with special initial conditions. For
more general cyclic solutions encountered in the above cases, the
nonadiabatic geometric phase contains in general an extra term in
addition to the one proportional to the classical Hannay angle.
Several examples are studied where the Hamiltonians are relatively
simple and analytic solutions are available. In these examples the
existence of cyclic solutions are discussed in detail. They exhibit
many possibilities: (1) cyclic solutions are available for all
$\tau$; (2) available for some $\tau$ and such $\tau$ may be found in
any interval $(\tau_0,+\infty)$ with $\tau_0>0$; (3) available for
some $\tau$ but such $\tau$ exists only in some finite interval
$(0,\tau_0)$; (4) not available at all. From the point of view of
wave packets, the motion in these examples also exhibits various
patterns. For a Hamiltonian of a definite nature, say, elliptic,
several different patterns of motion are possible, depending on the
value of some parameter in the Hamiltonian. There exists some
critical value of the parameter at which some kind of phase
transition happens. When the parameter goes through it, the motion
changes from one pattern to another, and at the critical value itself
the motion has an independent pattern.

\begin{acknowledgments}
This work was supported by the National Natural Science Foundation of
the People's Republic of China (Grant number: 10275098), and by the
Foundation of the Advanced Research Center of Sun Yat-Sen University
(Grant number: 02P3).
\end{acknowledgments}

\appendix*
\section{}

Our approach to the solution of the Schr\"odinger equation with
time-dependent Hamiltonians was first used in our previous paper
\cite{jpa03} to the system of general spin moving in an arbitrarily
varying magnetic fields, and is further developed in this paper. The
Hamiltonian for the spin in a magnetic field is an element of the
SO(3) algebra while that for the time-dependent harmonic oscillator
is one of the SO(2,1) algebra. In this appendix we briefly discuss
how to extend the formalism to a system where the Hamiltonian is an
element of a more general Lie algebra.

Consider a Lie algebra with the generators $K_a$ ($a=1,2,\ldots, l$),
satisfying the commutation relation
\begin{equation}\label{a1}
[K_a, K_b]=\ri f_{ab}{}^c K_c,
\end{equation}
where $f_{ab}{}^c$ are the structure constants. Define
\begin{equation}\label{a2}
g_{ab}=f_{ac}{}^d f_{bd}{}^c=\mathrm{tr}(f_a f_b),
\end{equation}
where $f_a$ is a matrix whose element is defined by
$(f_a)_b{}^c=f_{ab}{}^c$. It is obviously symmetric: $g_{ba}=g_{ab}$.
For a semisimple Lie algebra, the matrix $g_{ab}$ is not singular,
and thus invertible. Its inverse matrix is denoted by $g^{ab}$,
satisfying $g^{ac}g_{cb}= \delta^a{}_b$. The matrices $g_{ab}$ and
$g^{ab}$ acts like metric tensors and can be used to lowering and
raising vector indices. For the SO(2,1) algebra with the generators
given in Eq. (\ref{2}), one should find
$g_{ab}=8\,\mathrm{diag}\,(1,1,-1)$ and $g^{ab}=(1/8)\,
\mathrm{diag}\,(1,1,-1)$. Thus the metric used in the main text is
different from this by a constant factor.

For simplicity we consider a semisimple Lie algebra of rank 2, say,
the SU(3) algebra. It contains two mutually commuting operators which
generates the Cartan subalgebra. Thus there are two Casimir
operators, given by
\begin{equation}\label{a3}
C_2=g^{ab}K_a K_b,\quad C_3=h^{abc}K_a K_b K_c,
\end{equation}
where $h^{abc}$ is obtained by raising the indices of $h_{abc}$, and
the latter is defined by
\begin{equation}\label{a4}
h_{abc}=\mathrm{tr}(f_a f_b f_c).
\end{equation}

Now consider a physical system whose Hamiltonian is an element of the
above semisimple Lie algebra. More specifically,
\begin{equation}\label{a5}
H(t)=\hbar \omega^a(t) K_a,
\end{equation}
where $\omega^a(t)$ are time-dependent frequency parameters, and the
generators $K_a$ are dimensionless. We define a $l$-component vector
$\bm u(t)=\bm(u_1(t),u_2(t),\ldots, u_l(t)\bm)$ by
\begin{equation}\label{a6}
u_a(t)=\langle K_a\rangle\equiv\bm(\psi(t), K_a\psi(t)\bm),
\end{equation}
where $\psi(t)$ is an arbitrary state of the system, that is, a
solution to the Schr\"odinger equation. It is easy to show that it
satisfies the equation
\begin{equation}\label{a7}
\dot u_a(t)=f_{ab}{}^c\omega^b(t) u_c(t).
\end{equation}
This is a system of $l$ linear differential equations. Now we define
a $l$-component vector $\bm e(t)=\bm(e_1(t),e_2(t),\ldots,
e_l(t)\bm)$ by the same equation, that is
\begin{equation}\label{a8}
\dot e_a(t)=f_{ab}{}^c\omega^b(t) e_c(t),
\end{equation}
and a nontrivial (nonzero) initial condition. It is then
straightforward to show that the two operators linear in $K_a$,
defined by
\begin{equation}\label{a9}
L_1(t)=g^{ab}e_a(t) K_b,\quad L_2(t)=h^{abc}e_a(t) e_b(t) K_c
\end{equation}
are invariant operators, and they commute with each other. Therefore,
the invariant operators can be obtained by solving a linear
differential equation for $\bm e(t)$. It can be similarly shown that
$g^{ab}e_a(t) e_b(t)$ and $h^{abc}e_a(t) e_b(t) e_c(t)$ are also
time-independent quantities. Thus among the $l$ components of $\bm
e(t)$ only $l-2$ are independent variables. We can parameterize $\bm
e(t)$ in some way similar to that in Eq. (\ref{15}). The independent
parameters will be denoted by $\xi(t)=\bm(\xi_1(t), \xi_2(t), \ldots,
\xi_{l-2}(t)\bm)$.

The subsequent steps in obtaining the time evolution operator depend
on the details of the Lie algebra. We can give only a rough sketch
here.

The crucial step is to find a unitary operator $Q(t)=Q\bm(\xi(t)\bm)$
such that
\begin{equation}\label{a10}
L_1(t)=Q(t) H_1 Q^\dag(t), \quad L_2(t)=Q(t) H_2 Q^\dag(t),
\end{equation}
where $H_1$ and $H_2$ are generators of the Cartan subalgebra. It
might be difficult to prove the existence of $Q(t)$ in the general
case. For a specific Lie algebra, however, one can try to find it by
practical calculations like those in Sec. \ref{s2}. Since $H_1$ and
$H_2$ commutes with each other, they have a complete set of common
eigenstates. These will be denoted by $\{\psi_{nm}\}$, satisfying
\begin{equation}\label{a11}
H_1\psi_{nm}=\lambda_n\psi_{nm},\quad H_2\psi_{nm}=\mu_m\psi_{nm}.
\end{equation}
Obviously, $\lambda_n$ and $\mu_m$ are also the eigenvalues of
$L_1(t)$ and $L_2(t)$, respectively.

Next we will show that if the initial state $\psi(0)$ of the system
satisfies
\begin{equation}\label{a12}
L_1(0)\psi(0)=\lambda_n\psi(0),\quad L_2(0)\psi(0)=\mu_m\psi(0),
\end{equation}
then the state $\psi(t)$ at later times will satisfy
\begin{equation}\label{a13}
L_1(t)\psi(t)=\lambda_n\psi(t),\quad L_2(t)\psi(t)=\mu_m\psi(t).
\end{equation}
We use here a different proof from the induction one employed in Sec.
\ref{s3}. It is easy to show that $\psi^{(1)}(t)\equiv [L_1(t)-
\lambda_n]\psi(t)$ satisfies
\begin{equation}\label{a14}
\ri\hbar\partial_t \psi^{(1)}(t)=H(t)\psi^{(1)}(t),
\end{equation}
where the equation $\ri\hbar\partial_t L_1(t)+[L_1(t),H(t)]=0$ has
been used. Consequently, $\psi^{(1)}(t)=U(t)\psi^{(1)}(0)$ where
$U(t)$ is the time evolution operator of the Schr\"odinger equation.
Because Eq. (\ref{a12}) leads to $\psi^{(1)}(0)=0$, we have
$\psi^{(1)}(t)=0$, and thus the first equation in (\ref{a13}) is
obtained. The second one can be proven in a similar way.

On account of the above results, we have
\begin{equation}\label{a15}
\psi(t)=\exp[\ri\alpha_{nm}(t)]Q(t)\psi_{nm},
\end{equation}
where $\alpha_{nm}(t)$ is a phase that cannot be determined by the
eigenvalue equation. However, it is not arbitrary. By the requirement
that $\psi(t)$ satisfies the Schr\"odinger equation, it can be
determined in terms of $\xi(t)$. Finally, one should manage to
replace $\lambda_n$ appearing in $\alpha_{nm}(t)$ and
$\alpha_{nm}(0)$ by $H_1$, and $\mu_m$ by $H_2$, and obtain
\begin{equation}\label{a16}
\psi(t)=Q(t)\exp[\ri\alpha(H_1, H_2, t)]Q^\dag(0)\psi(0).
\end{equation}
Because all the operators in the above equation are independent of
$n$ and $m$, and because the set $\{\psi_{nm}\}$ is complete, we
obtain the time evolution operator
\begin{equation}\label{a17}
U(t)=Q(t)\exp[\ri\alpha(H_1, H_2, t)]Q^\dag(0).
\end{equation}
The time dependence in both $Q(t)$ and $\alpha(H_1, H_2, t)$ comes
from $\xi(t)$ (and $\omega^a(t)$ of course). Therefore, the time
evolution operator is also obtained by solving the linear different
equation for $\bm e(t)$.



\begin{thebibliography}{99}
\baselineskip 15pt

\bibitem{natur}F. Hertweck and A. Schl\"uter, Z. Naturforsch {\bf
12A}, 844 (1957).

\bibitem{paul}W. Paul, Rev. Mod. Phys. {\bf 62}, 531 (1990).

\bibitem{lewis1}H. R. Lewis, Jr., Phys. Rev. Lett. {\bf 18},
510 (1967); J. Math. Phys. {\bf 9}, 1976 (1968).

\bibitem{lewis2}H. R. Lewis and W. B. Riesenfeld,
J. Math. Phys. {\bf 10}, 1458 (1969).

\bibitem{gerry}C. C. Gerry, Phys. Rev. A {\bf 35}, 2146 (1987).

\bibitem{selez}A. N. Seleznyova, Phys. Rev. A {\bf 51},
950 (1995), and references therein.

\bibitem{berry}M. V. Berry, Proc. R. Soc. Lond. A {\bf 392}, 45
(1984).

\bibitem{simon}B. Simon, Phys. Rev. Lett. {\bf 51}, 2167 (1983).

\bibitem{aha}Y. Aharonov and J. Anandan, Phys. Rev. Lett. {\bf 58},
1593 (1987).

\bibitem{sam}J. Samuel and R. Bhandari, Phys. Rev. Lett. {\bf 60},
2339 (1988).

\bibitem{wu-li}Y.-S. Wu and H.-Z. Li, Phys. Rev. B {\bf 38}, 11907
(1988).

\bibitem{jordan}T. F. Jordan, Phys. Rev. A {\bf 38}, 1590 (1988).

\bibitem{resource}J. Anandan, J. Christian and K. Wanelik, Am. J.
Phys. {\bf 65}, 180 (1997).

\bibitem{li-book}H.-Z. Li, {\it Global Properties of Simple Physical
Systems--Berry's Phase and Others} (Shanghai Scientific \& Technical,
Shanghai, 1998) (in Chinese).

\bibitem{chat}S. Chaturvedi, M. S. Sriram and V. Srinivasan,
J. Phys. A {\bf 20}, L1071 (1987).

\bibitem{wang-sj}F.-L. Li, S.-J. Wang, A. Weiguny and D. L. Lin,
J. Phys. A {\bf 27}, 985 (1994).

\bibitem{ji}J.-Y. Ji, J. K. Kim, S. Y. Kim and K. S. Soh,
Phys. Rev. A {\bf 52}, 3352 (1995).

\bibitem{lewis3}H. R. Lewis, W. E. Lawrence and J. D. Harris,
Phys. Rev. Lett. {\bf 77}, 5157 (1996).

\bibitem{ge}Y.-C. Ge and M. S. Child, Phys. Rev. Lett. {\bf 78},
2507 (1997); Phys. Rev. A {\bf 58}, 872 (1998).

\bibitem{liu}J. Liu, B. Hu and B. Li, Phys. Rev. Lett. {\bf 81},
1749 (1998).

\bibitem{wang-xb}X.-B. Wang, L. C. Kwek and C. H. Oh,
Phys. Rev. A {\bf 62}, 032105 (2000).

\bibitem{fuent}I. Fuentes-Guridi, S. Bose and V. Vedral,
Phys. Rev. Lett. {\bf 85}, 5018 (2000).

\bibitem{wang-pra}S.-J. Wang, Phys. Rev. A {\bf 42}, 5107 (1990).

\bibitem{wagh1}A. G. Wagh and V. C. Rakhecha, Phys. Lett. A {\bf
170}, 71 (1992).

\bibitem{wagh2}A. G. Wagh and V. C. Rakhecha, Phys. Rev. A {\bf 48},
R1729 (1993).

\bibitem{fer}D. J. Fern\'andez C., L. M. Nieto, M. A. del Olmo and
M. Santander, J. Phys. A {\bf 25}, 5151 (1992).

\bibitem{fer-pla}D. J. Fern\'andez C. and O. Rosas-Ortiz,
Phys. Lett. A {\bf 236}, 275 (1997).

\bibitem{layton}E. Layton, Y. Huang and S-I Chu, Phys. Rev. A {\bf
41}, 42 (1990).

\bibitem{gao}X.-C. Gao, J.-B. Xu and T.-Z. Qian, Phys. Lett. A {\bf
152}, 449 (1991).

\bibitem{ni}G.-J. Ni, S.-Q. Chen, and Y.-L. Shen, Phys. Lett. A
{\bf 197}, 100 (1995).

\bibitem{zhang}Y.-D. Zhang, G. Badurek, H. Rauch and J. Summhammer,
Phys. Lett. A {\bf 188}, 225 (1994).

\bibitem{zhu00}S.-L. Zhu, Z. D. Wang and Y.-D. Zhang, Phys. Rev. B {\bf
61}, 1142 (2000).

\bibitem{pra01}Q.-G. Lin, Phys. Rev. A {\bf 63}, 012108 (2001).

\bibitem{jpa01}Q.-G. Lin, J. Phys. A {\bf 34}, 1903 (2001).

\bibitem{jpa02}Q.-G. Lin, J. Phys. A {\bf 35}, 377 (2002).

\bibitem{jpa03}Q.-G. Lin, J. Phys. A {\bf 36}, 6799 (2003).

\bibitem{ni-book}G.-J. Ni and S.-Q. Chen, {\it Advanced Quantum
Mechanics} (Fudan Univ. Press, Shanghai, 2000) (in Chinese).

\bibitem{hannay1}J. H. Hannay, J. Phys. A {\bf 18}, 221 (1985).

\bibitem{berry1}M. V. Berry, J. Phys. A {\bf 18}, 15 (1985).

\bibitem{ber-han}M. V. Berry and J. H. Hannay, J. Phys. A {\bf 21},
L325 (1988).

\bibitem{pla02}Q.-G. Lin, Phys. Lett. A {\bf 298}, 67 (2002).

\bibitem{wilcox}R. M. Wilcox, J. Math. Phys. {\bf 8}, 962 (1967).

\end{thebibliography}
\end{document}